# The Issue with Special Issues: when Guest Editors Publish in Support of Self

Paolo Crosetto[1*], Pablo Gómez Barreiro[2], Mark Austin Hanson[3*]

1. Univ. Grenoble Alpes, INRAE, CNRS, Grenoble INP, GAEL; Grenoble, France
2. Department of Science Operations, Royal Botanic Gardens, Kew; Wakehurst, UK
3. Centre for Ecology and Conservation, University of Exeter; Penryn, Cornwall, UK

* corresponding authors – PC: paolo.crosetto@inrae.fr, MAH: m.hanson@exeter.ac.uk

Open Researcher and Contributor ID (ORCID)

PC: 0000-0002-9153-0159
PGB: 0000-0002-3140-3326
MAH: 0000-0002-6125-3672

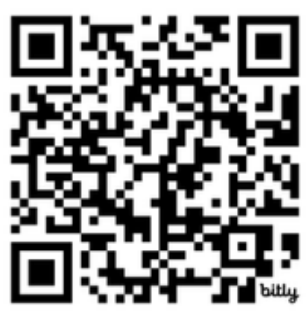

← Explore our data

## Abstract

The recent exceptional growth in special issues has led to the largest delegation of editorial power in the history of scientific publishing. Has this power been used responsibly? We provide the first systematic analysis of *endogeny*, the practice of publishing articles in one's own special issue. While moderate levels of endogeny are common, excessive endogeny constitutes scientific misconduct, as it stems from a clear conflict of interest. We define special issues containing more than 33% endogeny as "SI-hacked." We build a dataset of over 100,000 special issues published in 2015-2025 by five leading publishers. The large majority of guest editors engage in endogeny responsibly, if at all. Nonetheless, despite endogeny policies by publishers and indexers, SI-hacking is endemic. All journals heavily relying on special issues host SI-hacking; more than 1,000 hacked SIs are published each year, hosting tens of thousands of endogenous articles. Egregious SI-hacking is rare, editors exceeding endogeny thresholds mostly to the extent that publishers allow them to. This is not good news, as it reflects a widespread normalisation of guest editor conflicts of interests. Fortunately, SI-hacking can be solved by enforcing existing commonsense policies. We provide data and analyses needed for indexers and regulators to act.

## Introduction

Special issues are getting less special. Guest edited collections – i.e. issues of an academic journal that are run by a guest editor – have long existed. They were usually meant to celebrate the work of a conference, or were organized by researchers around a topical subject. While this traditional incarnation of special issues still exists, some publishers have reinvented their purpose, size, and scope, transforming them into one of the main engines behind the recent explosive growth in the number of scientific articles and strain on practicing scientists [1]. Indeed, the share of articles published in special issues, rather than regular issues, has increased dramatically (Table S1), particularly at for-profit gold Open Access (OA) publishers where special issue articles account for the vast majority of published papers – 69% for Frontiers Media (Frontiers) and 88% for Multi-Disciplinary Publishing Institute (MDPI) in 2022. Special has become the new normal.

This happened because special issues serve the historically recent alignment of incentives of open-access publishers with the pre-existing needs of authors under "publish or perish" [2,3]. Authors, pressured by a need to evidence productivity, look for easier and faster ways to publish their studies [4]. Special issues provide such an outlet, with generally lower turnaround times and higher article acceptance rates [1]. In the Gold OA model, where authors or institutions pay article processing charges (APCs), publishers can increase revenues by publishing more papers. Special issues shift part of the workload of publishing these articles onto volunteer editors, further increasing their profitability. Indeed, most guest editors work *pro bono* for the success of their special issue for the promise of prestige, the opportunity to write review articles (which receive more citations on average [5]), and the rewards that publishing articles offers themselves and their colleagues [6]. Guest editors also mobilize their networks to attract more papers and source reviewers, further increasing the publisher's revenue from APCs [7,8].

The impressive recent growth of special issues has resulted in the largest delegation of editorial power in the history of academic publishing. Hundreds of thousands of researchers have been offered the opportunity to (guest) edit, opening up a role that was traditionally reserved for selected leading experts. This can be positive, insofar as it combats systemic biases, making publishing inclusive and accessible to a more diverse pool of communities. But it also comes with a dilution of control over the published literature, which in turn entails potential costs and threats to research integrity. There is evidence that hosting many special issues is a hallmark of questionable journals that enable research paper mill activity[1] [9–12]. More generally, special issues run on delegated trust, and as such present a common pool resource problem [13]: common pool management relies on all users to access resources responsibly. However, when a resource can be accessed by a large number of people, even slight overuse by multiple individuals can cause depletion. In the case of Special Issues, each invited guest editor has an individual incentive to be lenient towards themselves or their network of connected authors. This leniency contributes to the erosion of the reputation of special issues as a whole, and even the broader public trust in scientific journals, which is founded upon rigorous and *unbiased* peer review. These problems balloon as the number of invited guest editors increases, especially if guest editor recruitment and special issue article solicitation is carried out through mass spam email campaigns [14] (and File S1). Whether the explosion of special issues is a net positive or negative for science depends, crucially, on whether the trust delegated to guest editors is used responsibly, or if it is used to disproportionately favour themselves and their colleagues compared to the usual publishing process.

In this paper we look in depth at one instance of such an abuse of delegated trust: *endogeny*, which we define as the practice of a guest editor publishing a non-editorial article in one's own special issue (per the Directory of Open Access Journals (DOAJ) [15]). While endogeny can be considered acceptable in some cases, poor academic practice in others, and even scientific misconduct in egregious cases, endogeny itself is not fraud. Indeed, to our knowledge, no publisher allows the direct conflict of interest of having guest editors handle their own articles. Nonetheless, endogeny *still* represents a conflict of interest, since publishers would be hard-pressed to reject papers submitted by guest editors, lest they lose their free labour and access to the editor's network of authors. Two things are clear: *First,* endogeny, independently of whether it is carried out in good or bad faith, benefits guest editors and helps them achieve their goals – be they validation, fame, visibility, recognition, or a larger CV. *Second,* egregious levels of endogeny are unanimously seen as abuse of the special issue model, implying that there must exist a threshold for how much guest editors can ignore their conflicts of interest to further their own goals, lest they stray into the realm of scientific misconduct.

Here we label special issues with very high rates of guest editor endogeny as SI-hacked, and the relative practice as SI-hacking (SI pronounced like "Psi"). SI-hacking is a practice that parallels *P*-hacking [16]: irrespective of its motivation – be it naive or intentional – it is poor academic practice. It is also akin to self-citation: systematic and excessive self-citation by authors, journals, or through citation cartel-like behaviour, is a signal of poor academic practice [1,9,17,18]. While acceptable levels of endogeny exist, the levels of endogeny we describe as SI-hacking stem from ignoring conflicts of interest, and objectively result in biased self-promotion beyond the norms of academic peer review. As such, the prevalence of SI-hacking in the literature is one metric to reveal whether the recent proliferation of special issues has been an overall net positive, or negative, for the scientific record.

[1] Research paper mills coordinate fraudulent article publishing in service to bad-faith authors.

In this article we build the largest existing dataset of special issue endogeny. This dataset is comprised of over one hundred thousand guest-edited article collections, containing just under one million papers published from 2015 to 2025. Using these data, we analyse the overall levels of endogeny, propose and defend a threshold to consider a special issue as SI-hacked, provide benchmarks that can be useful for scientific indexers in monitoring special issue academic practice, and reveal how much guest editor SI-hacking has affected the scientific record.

Overall, mean endogeny sits at about 1 paper in 7, and the vast majority of special issues are not hacked; this is welcome news. However, all journals following a special-issue-centered growth model host SI-hacking, for a total of over one thousand hacked special issues per year. We found many egregious cases of SI-hacking, with guest editors authoring >75% of the articles in their issue. However, most hacked special issues contain levels of endogeny that are perhaps best described as "poor practice."

Fortunately, putting a stop to SI-hacking is easy. Endogenous publishing is easily tracked with existing metadata, and can be squashed by simply enforcing the *already-existing* policies of indexers and publishers. Indeed, we provide tentative evidence as to the positive impact of indexer policies in this realm. This article presents both the data to assess compliance with policy, and a framework for how to collect and analyse such data moving forward.

## Methods and Data

### *Data overview and inclusion criteria*

We obtained data from all articles published over the years 2015 to 2025 in closed special issues (i.e., not accepting more papers at the time of obtaining the data), from a comprehensive list of journals published by MDPI, Frontiers, BioMed Central (BMC), the Springer Nature Discover journal series (Discover), and the UK Royal Society special issue-only journals Philosophical Transactions of the Royal Society A and B (Phil Trans). A summary of our dataset is provided in Table 1.

Our dataset includes major publishers that have used special issues as an engine for growth (MDPI, Frontiers), publishers that show increasing special issue totals in recent years (BMC, Discover), and also a small society publisher (Phil Trans), covering a broad spectrum of special issue publishing practices. We further included Discover given its recent uptick in special issues, and observations of its journal naming conventions and impact inflation mirroring those of MDPI [19]. We provide a summary of publisher special issue policies and practices in Table S2.

Our data cover both for-profit and non-profit publishers, including journals publishing from 10% to 100% of their articles in special issues, and releasing articles as they are accepted, or only simultaneously as a collection.

The information required of guest editors on initial contact with the publisher also varies markedly, with some publishers asking would-be guest editors only for a short (≤350 words) expression of interest (MDPI, Frontiers, Discover), while Phil Trans asks for a full ~5500 word outline justifying the suitability of the scientific conversation, the timing as topical given recent major innovations, and why the proposed articles should be published as a collection rather than independently. On the other hand, BMC special issues are conceived of and led by senior editors, and guest editors are secondarily recruited to help refine the concept and provide editorial topic expertise. On rare occasion within our dataset, BMC can also define their article collections post-hoc by pooling already-published articles, sometimes from across several years, into a collection; we removed most such post-hoc collections from our dataset (see below).

**Table 1: overview of data on special issues across publishers.**

| | | Totals | | | Special Issues by year | | | | | | | | | | |
|---|---|---|---|---|---|---|---|---|---|---|---|---|---|---|---|
| PUBLISHER | DATE SCRAPED | PAPERS | JOURNALS | SI | 2015 | 2016 | 2017 | 2018 | 2019 | 2020 | 2021 | 2022 | 2023 | 2024 | 2025* |
| Overall | — | 1,009,094 | 904 | 113,302 | 1,415 | 1,624 | 2,070 | 3,550 | 5,514 | 8,149 | 14,120 | 21,166 | 24,845 | 19,910 | 10,939 |
| MDPI | Sep '25 | 791,248 | 382 | 88,556 | 812 | 1,101 | 1,532 | 2,675 | 4,482 | 7,017 | 11,286 | 14,917 | 18,977 | 16,183 | 9,574 |
| Frontiers | Sep '25 | 186,384 | 175 | 21,682 | 450 | 385 | 408 | 740 | 898 | 953 | 2,642 | 6,054 | 5,615 | 2,903 | 634 |
| BMC | Mar-Jun '25 | 18,621 | 308 | 1,749 | 99 | 85 | 76 | 82 | 71 | 71 | 92 | 90 | 156 | 607 | 320 |
| Discover | Jun '25 | 5,886 | 46 | 795 | 2 | 1 | 2 | 1 | 11 | 56 | 48 | 53 | 47 | 169 | 405 |
| Phil Trans | Feb '25 | 6,955 | 2 | 520 | 52 | 52 | 52 | 52 | 52 | 52 | 52 | 52 | 50 | 48 | 6 |

*Incomplete data for 2025, only up to 9 months included.

Special issues also vary in size. Each publisher applies its own criteria for what constitutes a successful issue, and publishers have different enforcement rates for their own policies. Frontiers requires a special issue to publish at least four articles to be considered viable, while MDPI suggests at least ten articles[2]. However, we found many closed special issues with fewer articles than the publisher policy's stated minimum. The number of articles in each special issue in our dataset ranges from 1 to 299, with publisher-specific distributions (Figure S1).

Finally, publishers differ in whether they invite submissions independent of the guest editor. Thus, the recruitment of guest editors, and how much involvement guest editors or publishers have in conceptualizing the topic, and in recruiting authors, varies across publishers (Table S2).

### *Data collection*

Data collection methods included requests to publishers, manual curation from web pages, web scraping in R (using *rvest* and the *MDPIexploreR* packages [20,21]), and text mining from provided sources (Table S2). Web scraping took place between February and September 2025. Author data was cross-referenced using CrossRef [22]. For BMC, web page architecture was poorly standardized preventing effective web scraping. As a result, we manually curated 1,929 BMC article collections, recording guest editor names listed as part of the special issue (when present), or in Editorial articles (when present). To filter out post-hoc collections made up of papers that were initially published independently, we excluded BMC special issues whose articles bore publication dates spanning a range greater than 2 years. While we have performed our best effort in accurately annotating BMC data, we nonetheless recommend caution in extrapolating trends from BMC data given these challenges.

Processed datasets included annotations of article types (e.g. editorial, original research, review), authors of each paper, and guest editors. Editorial articles were excluded from special issue article totals and endogeny calculations. Name-cleaning and matching functions were adapted by publisher data type. Name cleaning involved removing titles, hyphens, extra spaces and converting character strings to lower case. Cleaned names were then reduced to combinations of given name and first surname, and then guest editor names were text-matched to author lists in R. If one or more guest editors were listed as an author, the article was flagged as endogenous. We will note that any errors arising from our methodology would most likely be false negatives, potentially underestimating the extent of reported endogeny and PISS behavior

[2] Strangely, these publishers and others (e.g. Elsevier) have policies in place to post-hoc change the status of an article from being part of a special issue to being a "regular" article in the event a special issue does not meet their minimum total articles (File S2). This distortion of the article's processing history may lead to an undercount of endogeny, as this process obscures the editorial context in which some articles were handled.

**Table 2: definitions of metrics used in this article.**

**Definitions of metrics used in this article**

| METRIC | APPLIES TO... | TYPE | DEFINITION |
|---|---|---|---|
| *Endogenous/Endogeny* | Paper | Binary | A binary definition denoting whether an author of the article is guest editor of the special issue it appears in. |
| *Endogeny volume* | Special Issue | Continuous | Share of endogenous articles in a special issue. |
| *SI-hacked* | Special Issue | Binary | A binary definition of whether a special issue has more than 1 in 3 articles that are endogenous |
| *Journal SI-hacking rate* | Journal | Continuous | Share of special issues classified as SI-hacked in a given journal |

The term endogeny has been used in various ways in previous literature [6,23,24]. To clarify our analysis, we provide definitions of the terms we use in this article in Table 2: We label a special issue article authored by at least one guest editor of the special issue, other than an editorial, as *endogenous*.

We define a special issue where more than 33% of articles (excluding editorials) are endogenous as SI-hacked. This one-in-three rule that we impose as a threshold for malpractice is debatable. The practice and tolerance of endogeny can vary across editorial boards, disciplines and publishers. Indeed, publishing an article in one's own special issue can be acceptable. Guest editors are usually experts in their field, and if they are convening a special issue, it may be expected that they have thoughts on the topic at hand. It is therefore quite common for them to publish an article in their own special issue. However, publishing in one's own special issue clearly entails a conflict of interest, and endogeny is a practice that scales badly: very high levels of endogeny are an exercise in academic narcissism, and have been considered a form of academic misconduct [6,15]. We reason that one guest editor should, at a minimum, elevate the voices of at least two other unrelated research groups if a special issue is to promote a broader conversation. Finally, the 1 in 3 rule allows us to better compare the many small special issues in our dataset with larger special issues, as integer rounding at low article totals means one or two endogenous papers can exceed the 25% or 30% values used by the DOAJ or Frontiers; indeed 27% of special issues in our dataset feature five papers or less (Figure S1).

Our endogeny threshold for demarcating SI-hacking does not directly align to any existing publisher or journal policies on allowable endogeny. While some journals do not mention specific limits, most publishers do. Existing thresholds can range from strict zero-tolerance policies (e.g., The Computer Journal [25]), to proportions like the current 25% threshold set by DOAJ (and used by Wiley [26], MDPI [FileS2], and Springer Nature [FileS2], which includes BMC and Discover), or 30% used by Frontiers [27]. They can alternately define endogeny in absolute number of articles (e.g., Phil Trans limits guest editors to author a maximum of three non-editorial articles, with a maximum of one as first/senior author [28]). The chosen 33% threshold (without considering editorial type articles) is therefore *higher* than all stated policies of major publishers, and we believe this threshold would be at least borderline acceptable to the sensibilities of most academics.

Finally, we also use "Journal SI-hacking rate" to refer to the proportion of special issues in a given journal that are classified as SI-hacked.

## Results

### *SI-hacking is concentrated in journals publishing many special issues*

The most relevant unit of analysis for endogeny and SI-hacking is a journal, as their editorial boards can welcome a lower or higher number of special issues independently of the publisher, and easily impose limits on endogeny if desired.

Of the 904 journals covered in our sample, 142 display no endogeny, and a further 302 only publish special issues with endogeny levels up to the 33% SI-hacking threshold (Figure 1 A, B). That is to say, half of the journals are immune to the problem as we define it. These journals, however, published only 2,527 of the 113,302 special issues in our dataset – between 2015-2024, years for which we have complete data, this was an average of about one special issue every two years. On the other hand, slightly over half of the journals in our dataset have published at least one hacked special issue (460 journals) averaging ~22 special issues per year. Thus, the journals without SI-hacking are, by and large, journals where special issues are – dare we say – special. It is the journals that use special issues as an engine for growth where guest editors regularly indulge in SI-hacking.

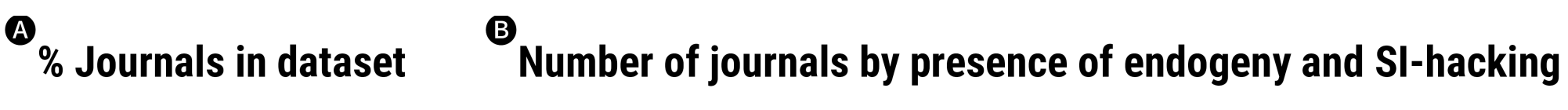

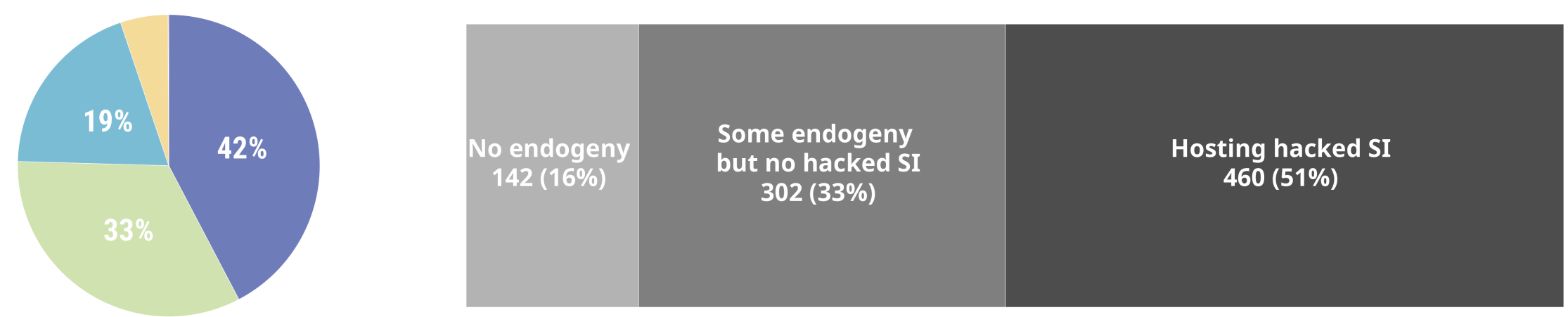

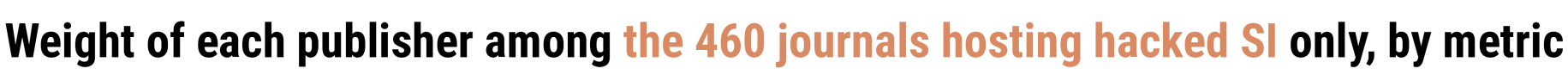

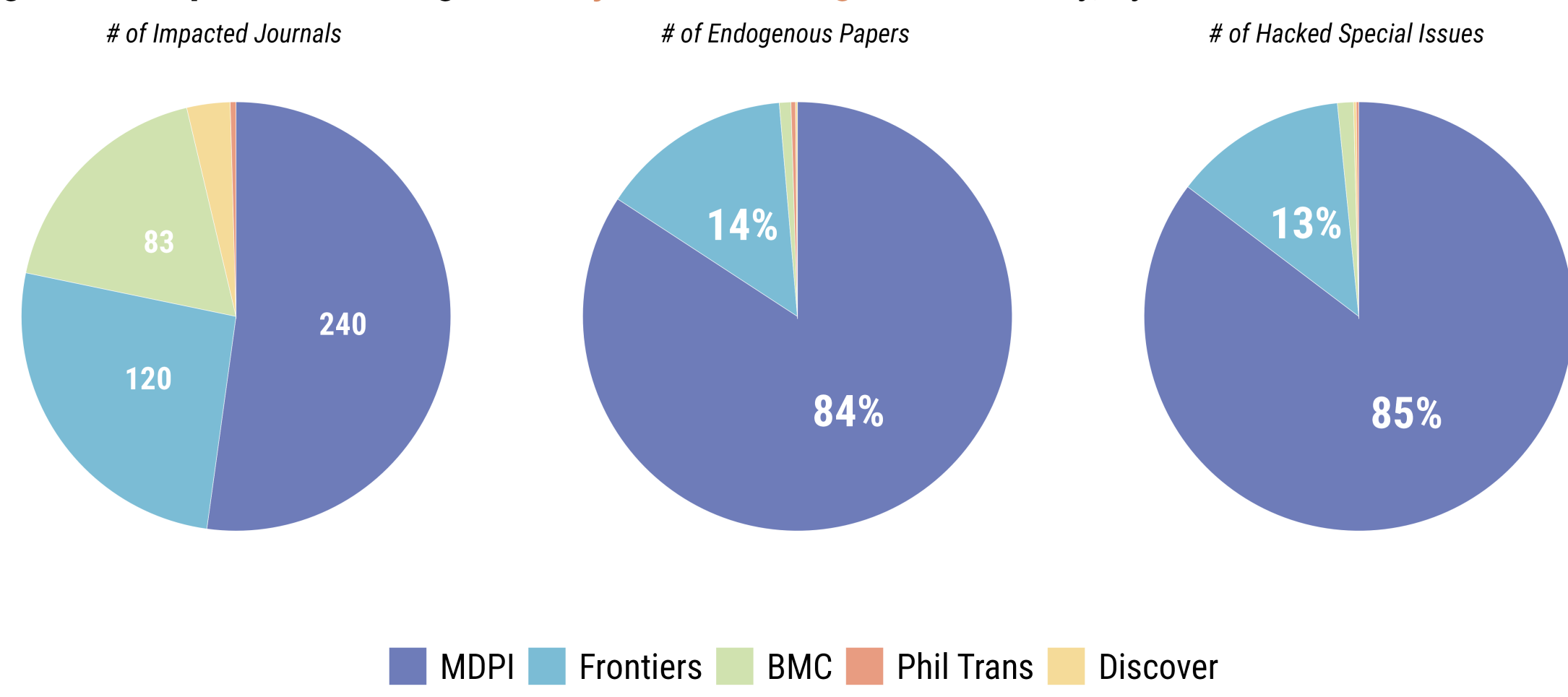

**Figure 1: prevalence of endogeny across journals and role of publishers**

We will next focus our analysis on only the 460 journals containing some hacked special issues. While our main focus is not on publishers, the overwhelming majority of SI-hacking took place in MDPI and Frontiers journals (Figure 1C). MDPI accounts for 42% of the journals hosting hacked special issues in our sample, but, by virtue of the sheer mass of articles published through the special issue format, MDPI published 84% of endogenous articles and 85% of special issues labeled as SI-hacked in our dataset. Frontiers accounts for the bulk of the remaining endogenous papers and hacked special issues. In absolute numbers the endogeny and SI-hacking problems highlighted here would not exist without, most specifically, MDPI.

### *Endogeny is endemic, but responsive to policies*

Of the 984,570 special issue articles published by journals hosting hacked special issues in our dataset, 113,302 are endogenous, resulting in an overall mean endogeny of 14.3% – i.e. one in seven special issue papers (excluding editorials) are authored by the guest editors. Endogeny rates within journals vary across publishers and years. The two journals published by Phil Trans display the highest mean endogeny (~19%), and these two journals have the most lenient endogeny policies (Table S2); this endogeny trend is driven about equally

**Table 3: endogeny levels and total endogenous papers by year and by publisher.**

| | Overall | | | | % Endogeny by year *(total endogenous papers)* | | | | | | | | | | |
|---|---|---|---|---|---|---|---|---|---|---|---|---|---|---|---|
| | # SI | # PAPERS | # ENDOGENOUS | % ENDOGENY | 2015 | 2016 | 2017 | 2018 | 2019 | 2020 | 2021 | 2022 | 2023 | 2024 | 2025* |
| Overall | 110,775 | 984,570 | 128,118 | **14.3 ± 0.1** | **10.3** *(1464)* | **11.9** *(1899)* | **12** *(2502)* | **12.8** *(4623)* | **13.9** *(7558)* | **15.8** *(12172)* | **17** *(21066)* | **15.1** *(27200)* | **14.3** *(27473)* | **12.3** *(14731)* | **13** *(7430)* |
| MDPI | 87,632 | 783,294 | 107,614 | **15.3 ± 0.1** | **8.7** *(758)* | **10.2** *(1135)* | **11** *(1695)* | **12.2** *(3335)* | **13.5** *(5966)* | **16.1** *(10596)* | **18.3** *(17916)* | **17.5** *(22546)* | **16.1** *(23993)* | **13.1** *(12744)* | **13.5** *(6930)* |
| Frontiers | 21,302 | 183,743 | 18,308 | **10.1 ± 0.2** | **12.4** *(554)* | **14.4** *(571)* | **15** *(656)* | **14.5** *(1100)* | **14.9** *(1338)* | **14.3** *(1358)* | **11.5** *(2908)* | **9.3** *(4466)* | **8.3** *(3274)* | **8.5** *(1706)* | **8.8** *(377)* |
| BMC | 730 | 6,261 | 734 | **15.6 ± 1.8** | **13.7** *(55)* | **26.6** *(66)* | **18.7** *(50)* | **19.7** *(59)* | **26.6** *(100)* | **16.8** *(43)* | **19.1** *(59)* | **13.9** *(50)* | **16.9** *(66)* | **10.6** *(139)* | **12.4** *(47)* |
| Discover | 591 | 4,317 | 142 | **4.7 ± 1.2** | **6.2** *(2)* | **1.8** *(1)* | **0** *(0)* | **0** *(0)* | **0** *(0)* | **2.9** *(18)* | **5.4** *(14)* | **5.8** *(17)* | **6.5** *(6)* | **4.8** *(23)* | **4.7** *(61)* |
| Phil Trans | 520 | 6,955 | 1,320 | **18.7 ± 1.2** | **14.6** *(95)* | **17.4** *(126)* | **15.1** *(101)* | **18.1** *(129)* | **20.9** *(154)* | **21.5** *(157)* | **24.7** *(169)* | **17.4** *(121)* | **19.3** *(134)* | **18.4** *(119)* | **18.1** *(15)* |

*Incomplete data for 2025, only up to 9 months included.

by both Phil Trans journals, despite these journals differing in their rate of hacked special issues (mean endogeny 18% and 19%, and SI-hacking rate 14% and 9% for PTA and PTB respectively, Table S3). For MDPI, Frontiers, and BMC, mean endogeny was between 10-15%. Broadly speaking, these major publishers indicate that endogeny comes naturally with special issues. On the other hand, the newer Discover journals series had a low rate of just ~5%, suggesting it has launched its multiple journals while keeping endogeny in check. Additionally, endogeny peaked in 2020-2021, and has slightly declined since, with 2024 and 2025 means lower than their peak values (Table 3, and Figure S2). In particular, year-over-year declines in MDPI (~3%) and BMC (~6%) between 2023 and 2024 may reflect a publisher- or journal-level policing response to DOAJ guidelines, updated in late 2023, which threatened to delist journals that host special issues containing more than 25% endogeny [15].

Overall, the average ratio of endogeny does not seem to be related to the business model of the publisher: the for-profit publisher Frontiers displays much lower levels of endogeny than for-profit BMC and MDPI. The total number of special issues is also not a good predictor of mean endogeny. Emphasized by the Phil Trans journals, journal-level mean endogeny is not necessarily a good indicator of SI-hacking rate. Instead, appreciable rates of endogeny seem to be a ubiquitous feature of journals publishing hacked special issues.

### *SI-hacking is concentrated in some journals*

The issue of SI-hacking is about concentration of endogenous articles in a given special issue. While measuring mean endogeny at the publisher level can give an overall view, this analysis obscures patterns of SI-hacking. While no publisher exceeded 20% mean endogeny, endogeny levels are highly heterogeneous across journals and special issues (Table 4).

We recovered >200,000 guest editors in our dataset, of which ~24,000 engaged in some level of SI-hacking. Table 4 shows the number and share of hacked special issues, breaking the total special issues into bins with different levels of endogeny. Our analysis reveals three main facts:

*First,* even within the set of journals hosting hacked special issues, the majority of special issues have a level of endogeny below the SI-hacking threshold. Across all publishers, at least 3 in 4 special issues (BMC, Phil Trans, MDPI) and more than 9 in 10 (Frontiers, Discover) are free from SI-hacking. Thus, the *vast* majority of guest editors take their role seriously, recognize and avoid the conflict of interest implicit with their role, and engage in endogeny responsibly, if at all.

**Table 4: Number and share of special issue by endogeny level**.

| | Acceptable SI | | | | Hacked SI | | | | | |
|---|---|---|---|---|---|---|---|---|---|---|
| | No Endogeny (0%) | | Responsible (up to 33%) | | Moderate (33 to 50%) | | Severe (50 to 75%) | | Egregious (75 to 100%) | |
| | N | % | N | % | N | % | N | % | N | % |
| Overall | 42,007 | 38% | 54,463 | 49% | 10,856 | 10% | 2,119 | 2% | 1,330 | 1% |
| MDPI | 29,513 | 33.7% | 45,927 | 52.4% | 9,181 | 10.5% | 1,781 | 2.0% | 1,230 | 1.4% |
| Frontiers | 11,531 | 54.1% | 7,929 | 37.2% | 1,484 | 7.0% | 294 | 1.4% | 64 | 0.3% |
| BMC | 400 | 54.8% | 167 | 22.9% | 108 | 14.8% | 29 | 4.0% | 26 | 3.6% |
| Phil Trans | 75 | 14.4% | 364 | 70.0% | 66 | 12.7% | 15 | 2.9% | – | – |
| Discover | 488 | 82.6% | 76 | 12.9% | 17 | 2.9% | – | – | 10 | 1.7% |

*Second,* a non-negligible minority of special issues can nonetheless be labeled as SI-hacked. This varies from 5% at Discover, to 9% at Frontiers, 14% at MDPI, 16% at Phil Trans, and 22% at BMC.

*Third,* the absolute magnitude of the phenomenon is striking. While only a minority of special issues are SI-hacked, approximately 1,300 hacked special issues are published per year across our dataset – most containing multiple articles. When looking at absolute numbers, the role of MDPI cannot be understated: over 11 years MDPI produced 12,192 hacked special issues; a total of ~1,110 per year. Frontiers published 1,842 or ~167 per year, while BMC, Discover and Phil Trans, who *collectively* host ~38% of the journals in our dataset, published just ~25 hacked special issues per year.

Finally, while ~121 special issues per year have a level of endogeny higher than 75%, i.e. at least 3 in 4 papers are authored by the guest editors, the bulk of SI-hacking occurs in special issues with levels better described as poor practice rather than egregious SI-hacking. This is only partially good news: as in most common pool resource management problems [13], the real danger does not come from rare, serious offenders, but from normalizing small-scale neglect of conflicts of interest by guest editors. [13]

***SI-hacking is prevalent at journals hosting many special issues***

Next we focus on the prevalence of SI-hacking across journals, as journal edtiorial boards are the relevant actors where control and regulation can happen.

Including all journals in our dataset, some broader patterns emerge in Figure 2. First, journals at either end of the spectrum – journals with no hacked special issues, or journals almost exclusively publishing hacked special issues – tend to publish very few special issues (typically ≤1 special issue every two years).

Second, journals with more than 1 in 2 hacked special issues are not only small, they are also rare – one at Frontiers, four at MDPI, and a dozen at BMC. However, journals that publish many special issues tend to have intermediate levels of SI-hacking. Again, this is not necessarily good news: while most researchers' and indexers' attention is monopolized by the small and rare serious offenders, the bulk of special issue hacking is published in lenient journals that normalize poor practice, even if they might block egregious attempts to SI-hack.

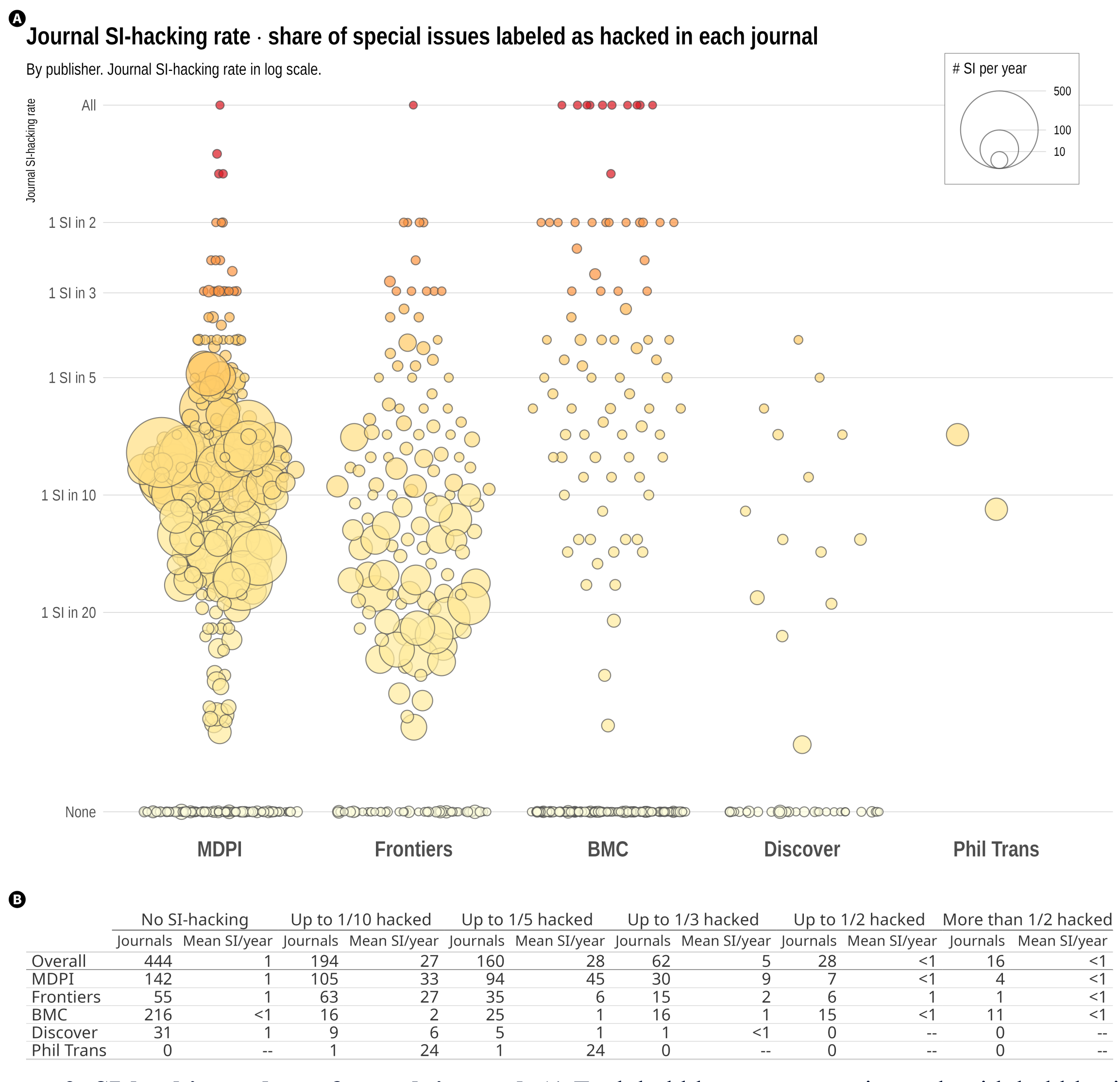

| | No SI-hacking | | Up to 1/10 hacked | | Up to 1/5 hacked | | Up to 1/3 hacked | | Up to 1/2 hacked | | More than 1/2 hacked | |
|---|---|---|---|---|---|---|---|---|---|---|---|---|
| | Journals | Mean SI/year | Journals | Mean SI/year | Journals | Mean SI/year | Journals | Mean SI/year | Journals | Mean SI/year | Journals | Mean SI/year |
| Overall | 444 | 1 | 194 | 27 | 160 | 28 | 62 | 5 | 28 | <1 | 16 | <1 |
| MDPI | 142 | 1 | 105 | 33 | 94 | 45 | 30 | 9 | 7 | <1 | 4 | <1 |
| Frontiers | 55 | 1 | 63 | 27 | 35 | 6 | 15 | 2 | 6 | 1 | 1 | <1 |
| BMC | 216 | <1 | 16 | 2 | 25 | 1 | 16 | 1 | 15 | <1 | 11 | <1 |
| Discover | 31 | 1 | 9 | 6 | 5 | 1 | 1 | <1 | 0 | -- | 0 | -- |
| Phil Trans | 0 | -- | 1 | 24 | 1 | 24 | 0 | -- | 0 | -- | 0 | -- |

**Figure 2: SI-hacking volume for each journal. A)** Each bubble represents a journal, with bubble size representing total special issues published between 2015 and 2025 by that journal. **B)** A numerical summary of the main features of the plot.

### *A few journals account for a majority of SI-hacking*

The vast majority of SI-hacking takes place in journals that normalize the poor practice of publishing special issues with endogeny beyond their own declared thresholds, but not necessarily to egregious levels. We can use our data to identify the journals that are most lenient towards this practice – i.e. journals with high concentrations of SI-hacking. Table 5 reports the 30 most lenient journals, filtering for journals that that have published at least 55 special issues in our dataset (at least 5 per year on average): among these, 24 are from MDPI, 5 from Frontiers, and 1 from Phil Trans.

These 30 most lenient journals account for 28% of the total hacked special issues in our full dataset of 904 journals. The 100 most lenient account for 69%, and the top 150 for 91%. These findings make it clear that action by publishers or editors-in-chief at a relatively small number of journals would solve most SI-hacking overnight.

In the Appendix, we report the complete statistics for all journals featuring SI-hacking.

**Table 5: 30 most lenient journals towards SI-hacking.** Here we rank journals by mean percent of hacked special issues. However, in absolute number of hacked special issues published, the most lenient journals are Sustainability (MDPI) and International Journal of Environmental Research and Public Health (MDPI). At the intersection, journals such as Water (MDPI) and Energies (MDPI) rank highly in both measures (see Table S7).

**Journals with highest journal SI-hacking rate**

Among all journals having published at least 5 SIs a year

| | | Endogeny at journal | | Special Issues | | |
|---|---|---|---|---|---|---|
| PUBLISHER | JOURNAL | MEAN | CONF. INT. | TOTAL | HACKED | % HACKED |
| **Frontiers** | *Frontiers in Built Environment* | 22% | [18% - 26%] | 122 | 30 | 25% |
| **MDPI** | *Journal of Marine Science and Engineering* | 22% | [20% - 23%] | 622 | 133 | 21% |
| **MDPI** | *Water* | 22% | [21% - 23%] | 1512 | 309 | 20% |
| **MDPI** | *Minerals* | 22% | [20% - 23%] | 633 | 127 | 20% |
| **MDPI** | *Medicina* | 22% | [20% - 24%] | 423 | 83 | 20% |
| **MDPI** | *Geosciences* | 18% | [15% - 21%] | 215 | 42 | 20% |
| **MDPI** | *Toxins* | 20% | [19% - 22%] | 400 | 75 | 19% |
| **MDPI** | *Infrastructures* | 17% | [13% - 22%] | 65 | 12 | 18% |
| **MDPI** | *Toxics* | 20% | [18% - 22%] | 235 | 43 | 18% |
| **MDPI** | *Actuators* | 17% | [14% - 20%] | 135 | 24 | 18% |
| **MDPI** | *Energies* | 20% | [19% - 20%] | 2915 | 485 | 17% |
| **MDPI** | *Metals* | 20% | [19% - 22%] | 784 | 126 | 16% |
| **MDPI** | *Atoms* | 18% | [13% - 23%] | 57 | 9 | 16% |
| **MDPI** | *Fishes* | 20% | [15% - 24%] | 114 | 18 | 16% |
| **Frontiers** | *Frontiers in Pain Research* | 15% | [10% - 20%] | 72 | 11 | 15% |
| **MDPI** | *Materials* | 19% | [18% - 19%] | 2539 | 378 | 15% |
| **MDPI** | *Brain Sciences* | 19% | [17% - 20%] | 513 | 76 | 15% |
| **Frontiers** | *Frontiers in Rehabilitation Sciences* | 15% | [10% - 19%] | 69 | 10 | 14% |
| **Phil Trans** | *PTA* | 18% | [16% - 20%] | 259 | 37 | 14% |
| **MDPI** | *Sports* | 19% | [14% - 23%] | 85 | 12 | 14% |
| **Frontiers** | *Frontiers in Marine Science* | 17% | [15% - 19%] | 462 | 65 | 14% |
| **MDPI** | *Cosmetics* | 18% | [12% - 24%] | 57 | 8 | 14% |
| **MDPI** | *Journal of Personalized Medicine* | 20% | [18% - 22%] | 374 | 52 | 14% |
| **MDPI** | *Computation* | 15% | [10% - 20%] | 72 | 10 | 14% |
| **Frontiers** | *Frontiers in Climate* | 15% | [10% - 19%] | 72 | 10 | 14% |
| **MDPI** | *Agronomy* | 18% | [16% - 19%] | 851 | 118 | 14% |
| **MDPI** | *Agriculture* | 18% | [17% - 20%] | 506 | 68 | 13% |
| **MDPI** | *Remote Sensing* | 16% | [15% - 17%] | 2058 | 275 | 13% |
| **MDPI** | *Foods* | 20% | [19% - 21%] | 1327 | 177 | 13% |
| **MDPI** | *Plants* | 18% | [17% - 19%] | 1188 | 153 | 13% |

We provide these data to help identify areas for action for indexers, publishers, and the interested reader (tables S3-S7).

## Discussion

The recent transformation of guest-edited collections from special to mundane has resulted in the largest delegation of editorial power in the history of scientific publishing. But with great delegation comes great responsibility. Despite all journals already being staffed by an editorial board, in our dataset we find that some ~200,000 persons were offered a guest editor role, with about 1 in 7 taking the opportunity to not only publish one, but several endogenous articles within their own special issue, a practice that we labeled SI-hacking. Even limited abuse of editorial power by guest editors can result in a deep systemic problem for the literature at large [13]. The scientific community clearly sees excessive endogeny as a form of academic misconduct, as publishers (Table S2) and policymakers have recommended limits [6,15]. Here, based on the largest dataset of endogeny in special issues to date, we present two key findings: **1)** the majority of guest editors use their power responsibly, **2)** a non-negligible number of them have been allowed to use their power as guest editors opportunistically, resulting in SI-hacking, despite publisher policies supposed to limit endogeny.

Low-level endogeny is endemic among guest editors wherever it is allowed. Indeed, the Phil Trans journals have the highest allowable endogeny (up to three articles per editor), and the highest mean endogeny among publishers. Where authors are invited to guest edit, many will use the opportunity to publish their own papers. Practiced responsibly, such endogeny can be tolerated. Thus, one might be tempted to dismiss SI-hacking as a niche and mostly innocuous problem. We argue otherwise.

First, the scope of SI-hacking in the literature, in terms of volume, results in the yearly publication of thousands of *excess* endogenous papers, i.e. *beyond the acceptable level of 33% per special issue*. This number is computed using our conservative definition of endogeny that includes only the guest editors themselves (per DOAJ [15]), a higher threshold to denote *SI-hacking* than any regulatory body (>33%), and a dataset of just 904 journals (Table 1). Were our definition of endogeny to include networks of guest editors *and* their colleagues (as in [6,24]), the total flagged articles would likely increase dramatically; this analysis is beyond the scope of the present study, but would be a worthwhile future endeavour given research paper mills coordinate sophisticated networks to mask such abuses [9].

Second, allowing hacked special issues to be published comes at considerable economic cost. Even with our conservative dataset and analysis, we identify 16,651 papers by guest editors published in their own special issues *beyond* the one-in-three allowance, and beyond the typical “one free article” that many publishers offer. If we consider all endogenous papers, the sum is 36,444. Assuming a conservative APC of €2000, €33-72 million has been spent over eleven years on hacked special issues. This amount could fund hundreds of research grants (e.g. through the French National Research Agency [29]). Guest editors offering easy routes to publishing for their colleagues (including citation cartels), suggests an even greater drain of scientific funding were we to use a more inclusive definition of endogeny (as in [6,24]). These sums, from just 904 journals, demonstrate how SI-hacking greatly drains time, money, and trust from the scientific enterprise [30,31].

Excessive endogeny is a form of academic misconduct. The industry leaders enabling this SI-hacking are undoubtedly MDPI and Frontiers through their scale of special issue hosting [1,32,33], but similar publishers (e.g. ACM, Springer-Nature, Elsevier, IEEE) have deployed special issues en masse much the same [11,34]. SI-hacking is also enabled by smaller publishers such as Phil Trans, and were enough small publishers to be aggregated, they could reach similar levels as industry giants like MDPI and Frontiers. SI-hacking would not be the massive problem it is without enablers among publishers.

Such publishers often offer the rebuttal that their journals provide a service for articles authors would write regardless [33,35]. But if this were true, then there is no reason to publish these papers through solicited special issues – such articles would be submitted for peer review through standard routes. Moreover, during our investigations we learned that many publishers have an open policy to modify DOI-indexed article metadata: practices at MDPI [36,37], Frontiers [FileS2], Elsevier [FileS2], and others intentionally remove the *"published as part of a special issue"* tag if the special issue as a whole fails to publish an arbitrary number of articles. This practice hides would-be endogeny, and so obscures the conflict of interest that the editors faced at the time of submission. Such behaviour has been termed "stealth correction", which is accurate, but perhaps falls short of communicating the seriousness of this distorition of scientific metadata. Our only reconciliation for how publishers can justify this practice in good faith is if they believe that the conflicts of interest intrinsic to endogenous special issue articles are not important for the reader, or regulatory bodies, to know. We profoundly disagree with such a sentiment. We call on academic indexers to immediately define policies that end this publisher practice.

Fortunately, unlike other challenges plaguing academic publishing, the solution to SI-hacking is simple. Preventing SI-hacking does not involve large collective action, nor reimagining the format of special issues; it only requires setting simple endogeny rules and enforcing them. Publishers and journals already have all the data and power to solve this problem – just check the author lists for guest editors and act. Indeed, our data show that endogeny policies work. For instance, DOAJ's 25% endogeny rule [17] seems to have had an impact on the overall levels of endogeny and SI-hacking given the drops observed from 2023 to 2024 (Table 3); publishers like Wiley, BMC and MDPI now have a 25% endogeny rule in their special issue policies ([27] and [FileS2]), and BMC mean endogeny dropped over ~6 percentage points in this period. When indexers throw their weight around, publishers seem to listen, and guest editors do fall in line.

Enforcing such policies requires good data. Our dataset is the fruit of a voluntary effort. It inevitably suffers from the noise that is added when our requests to publishers go unanswered, requiring us to web-scrape or perform painful manual curation. Efforts like ours, and the ones of scientific sleuths [38–40], cannot be relied upon continuously to feed indexers such as Clarivate or DOAJ the information they need to enforce their policies. We urge publishers to provide these data to open repositories such as OpenAlex [41] or CrossRef [22], including permanent version-specific identifiers needed to track endogeny (e.g., special issue, author, and editor metadata). Publication of these metadata transparently is not merely a suggestion, it is needed to combat academic misconduct. Journals failing to provide these data protect unscrupulous editors from policing efforts, and further betray the trust of honest guest editors.

## Conclusion

Here we have defined and framed the impact of SI-hacking in clear terms. We provide the data necessary for journals, publishers, and indexers to act.

Recent studies on scientific fraud and hidden conflicts of interest emphasize the scope of how authors can betray the trust that journals place in them [11,40]. Endogeny sits within a spectrum ranging from acceptable practice if done responsibly, to poor practice if done excessively, to egregious abuse that is best described as academic misconduct. Even within egregious abuse, there are likely cases where guest editors did not participate in SI-hacking with wrongful intent; nonetheless, the publication of these articles benefits the CV of the guest editors, and these articles are handled within a potentially biased review process suffering from the conflicts of interest intrinsic to the special issue model of publishing.

While addressing the forces behind the strain on scientific publishing are complex, SI-hacking relies on data that is transparently presented to the journal editors by requirement. As such, any event of egregious SI-hacking is a level of editorial negligence akin to accepting papers that overtly include AI slop [42]. We propose that SI-hacking rates can become a readily-available bibliometric litmus test for academic indexers when considering a journal for inclusion.

We also hope our analysis raises the long-overdue conversation on the less egregious, but still poor practice, of special issue endogenous publishing in general. While endogenous articles in hacked special issues are not necessarily compromised, we fail to see a *need* for authors intent on publishing multiple articles in a special issue to also guest edit for that special issue. This choice introduces many conflicts of interest to the authors, and to the journal, that are entirely unnecessary. Moreover, such practice opens the door to unscrupulous authors looking to SI-hack, publishing in support of self over the broader community. In response, would-be guest editors can normalize recusing themselves from guest-editing for a special issue entirely if they also intend to publish an article in that issue. Journals could also implement policies to restrict the number of guest editors per special issue, heading off the potential for numerous editor-author conflicts of interest within a special issue. Indeed, unlike other issues plaguing scientific publishing, *egregious* SI-hacking clearly requires both the guest editor(s) to engage in academic misconduct, and publishers to be negligent in their responsibility of overseeing the peer review process.

We propose indexers should not only enforce commonsense endogeny policies, but should also develop near-zero tolerance policies for egregious SI-hacking – not because it is the worst form of misconduct, but because it is extremely easy to detect. Failure to do so by journals cannot be reconciled with their role as gate keepers of the science publishing common resource pool.

## Acknowledgements

We thank eLife for providing special issue data (seven special issues, no SI-hacking) that were ultimately excluded from the article. We thank Dan Brockington for providing feedback on the manuscript prior to its public release. Preliminary findings of this manuscript were presented at the New Horizon conference in Nijmegen (The Netherlands), the Peer-Community In webinar, the Publish or Vanish seminar in Milan (Italy), the Saclay open Science Webinar, Paris (France), the University of Meknes (Morocco), the CIRAD Open science day, Montpellier (France), the University of Milan (Italy), and the Open Science day at the University of Brest (France). We thank conference, seminar and workshop participants and four anonymous referees for their comments.

### Author contributions

Web scraping was performed by PGB and PC; manual collection from BMC by MAH and PC. Data analysis in R was done by PC, PGB, and MAH. Conceptualisation was performed collectively by all authors, and all authors contributed to writing and revising the manuscript to produce the final article.

### Data availability

The replication package for our analysis, including the data and scripts to generate the figures and tables in this paper is available at https://doi.org/10.6084/m9.figshare.31045390.

We have also made our data available to explore through a Shiny App accessible at: https://paolocrosetto.shinyapps.io/Editors_as_authors/.

### Competing interests

The authors have no competing interests to declare.

### Funding

This work was a labour of love, and was not externally funded.

## References

1. Hanson MA, Barreiro PG, Crosetto P, Brockington D. The strain on scientific publishing. Quant Sci Stud. 2024; 1–29. doi:10.1162/qss_a_00327
2. Fanelli D. Do pressures to publish increase scientists' bias? An empirical support from US States Data. PloS One. 2010;5: e10271. doi:10.1371/journal.pone.0010271
3. Grimes DR, Bauch CT, Ioannidis JPA. Modelling science trustworthiness under publish or perish pressure. R Soc Open Sci. 2018;5: 171511. doi:10.1098/rsos.171511
4. Quan W, Chen B, Shu F. Publish or impoverish: An investigation of the monetary reward system of science in China (1999-2016). Aslib J Inf Manag. 2017;69: 486–502. doi:10.1108/AJIM-01-2017-0014
5. Miranda R, Garcia-Carpintero E. Overcitation and overrepresentation of review papers in the most cited papers. J Informetr. 2018;12: 1015–1030. doi:10.1016/j.joi.2018.08.006
6. COPE Council. COPE Guidelines: Guest edited collections. 2025. doi:10.24318/Bp64sd1c
7. Butler L-A, Matthias L, Simard M-A, Mongeon P, Haustein S. The oligopoly's shift to open access: How the big five academic publishers profit from article processing charges. Quant Sci Stud. 2023; 1–22. doi:10.1162/qss_a_00272
8. Shu F, Larivière V. The oligopoly of open access publishing. Scientometrics. 2024;129: 519–536. doi:10.1007/s11192-023-04876-2
9. Abalkina A. Publication and collaboration anomalies in academic papers originating from a paper mill: Evidence from a Russia-based paper mill. Learn Publ. 2023; leap.1574. doi:10.1002/leap.1574
10. Bishop DVM. Red flags for paper mills need to go beyond the level of individual articles: a case study of Hindawi special issues. PsyArXiv; 2023 Feb. doi:10.31234/osf.io/6mbgv
11. Richardson RAK, Hong SS, Byrne JA, Stoeger T, Amaral LAN. The entities enabling scientific fraud at scale are large, resilient, and growing rapidly. Proc Natl Acad Sci. 2025;122: e2420092122. doi:10.1073/pnas.2420092122
12. Zhuang H, Liang L, Acuna DE. Estimating the predictability of questionable open-access journals. Sci Adv. 2025;11: eadt2792. doi:10.1126/sciadv.adt2792
13. Ostrom E. Governing the commons: the evolution of institutions for collective action. 10th printing. Cambridge: Cambridge University Press; 2019.
14. Brockington D. MDPI Experience Survey Results. In: Dan Brockington - Research and Writings [Internet]. 18 Apr 2021 [cited 4 Dec 2025]. Available: https://danbrockington.com/2021/04/18/mdpi-experience-survey-results/
15. DOAJ. We have amended our Special Issues criteria. 25 Jan 2024 [cited 19 Oct 2025]. Available: https://blog.doaj.org/2024/01/25/weve-amended-our-special-issues-criteria/
16. Simonsohn U, Nelson LD, Simmons JP. P-curve: a key to the file-drawer. J Exp Psychol Gen. 2014;143: 534–547. doi:10.1037/a0033242
17. Chorus C, Waltman L. A Large-Scale Analysis of Impact Factor Biased Journal Self-Citations. Glanzel W, editor. PLOS ONE. 2016;11: e0161021. doi:10.1371/journal.pone.0161021
18. Wilhite A, Fong EA, Wilhite S. The influence of editorial decisions and the academic network on self-citations and journal impact factors. Res Policy. 2019;48: 1513–1522. doi:10.1016/j.respol.2019.03.003
19. Crosetto P, Hanson MA, Brockington D, Gómez Barreiro P. Springer Nature Discovers MDPI. In: The Daily Strain [Internet]. 15 June 2025 [cited 19 Oct 2025]. Available: https://hal.science/hal-05113253
20. Wickham H. rvest: Easily Harvest (Scrape) Web Pages. 2025. Available: https://rvest.tidyverse.org/
21. Gómez Barreiro P. MDPIexploreR: Web Scraping and Bibliometric Analysis of MDPI Journals. 2024. Available: https://github.com/pgomba/MDPI_exploreR
22. Hendricks G, Tkaczyk D, Lin J, Feeney P. Crossref: The sustainable source of community-owned scholarly metadata. Quant Sci Stud. 2020;1: 414–427. doi:10.1162/qss_a_00022
23. Nazarovets M, Nazarovets S. Editorial endogamy and endogeny: Challenges on the path to DOAJ. In: DOAJ blog [Internet]. 4 Dec 2025 [cited 5 Dec 2025].

Available:
https://blog.doaj.org/2025/12/04/editorial-endogamy-and-endogeny-challenges-on-the-path-to-doaj/
24. Sadowski I, Zamęcki Ł. Endogeny in measuring research excellence. In-house publishing and conflict of interests in Polish science evaluation. Res Policy. 2025;54: 105134. doi:10.1016/j.respol.2024.105134
25. The Computer Journal. Proposals for Special Issues. 2025 [cited 19 Oct 2025]. Available: https://academic.oup.com/comjnl/pages/special-issues-instructions
26. Wiley. Guest editor responsibilities and guidelines for Special Issues. In: Editor Resources [Internet]. 2025 [cited 19 Oct 2025]. Available: https://www.wiley.com/en-in/publish/article/editor-resources/special-issues/guidelines
27. Frontiers. Topic editor guidelines. In: Frontiers Host Editor Guidelines [Internet]. 2022 [cited 19 Oct 2025]. Available: https://www.frontiersin.org/Design/pdf/Host_Editor_Guidelines.pdf
28. The Royal Society. Information for Guest Editors. 2025 [cited 19 Oct 2025]. Available: https://royalsocietypublishing.org/rsta/guest-editors
29. Agence National de la Recherche. 2024 Annual Report. 2025. Available: https://anr.fr/fileadmin/documents/2025/ANR-Annual-Report-2024.pdf
30. Beigel F, Brockington D, Crosetto P, Derrick G, Fyfe A, Barreiro PG, et al. The Drain of Scientific Publishing. arXiv; 2025. doi:10.48550/ARXIV.2511.04820
31. Sabel B, Larhammar D. Reformation of science publishing: the Stockholm Declaration. R Soc Open Sci. 2025;12: 251805. doi:10.1098/rsos.251805
32. Oviedo-García MÁ. Journal citation reports and the definition of a predatory journal: The case of the Multidisciplinary Digital Publishing Institute (MDPI). Res Eval. 2021;30: 405–419a. doi:10.1093/reseval/rvab020
33. Grove J. Finland downgrades MDPI and Frontiers – will others follow suit? TImes Higher Education. 20 Dec 2024. Available: https://www.timeshighereducation.com/news/finland-downgrades-mdpi-and-frontiers-will-others-follow-suit. Accessed 7 Nov 2025.
34. Sandnes FE. Are there too many papers by the same authors within the same conference proceedings? Norms and extremities within the field of human–computer interaction. Scientometrics. 2025;130: 1659–1699. doi:10.1007/s11192-025-05271-9
35. Brainard J. Fast-growing open-access journals stripped of coveted impact factors. 2023. doi:10.1126/science.adi0098
36. Aquarius R, Schoeters F, Wise N, Glynn A, Cabanac G. The Existence of Stealth Corrections in Scientific Literature—A Threat to Scientific Integrity. Learn Publ. 2025;38: e1660. doi:10.1002/leap.1660
37. Bishop. Now you see it, now you don’t: the strange world of disappearing Special Issues at MDPI. In: BishopBlog [Internet]. 2024 [cited 19 Oct 2025]. Available: https://deevybee.blogspot.com/2024/09/now-you-see-it-now-you-dont-strange.html
38. O’Grady C. Image sleuth faces legal threats. Science. 2021;372: 1021–1022. doi:10.1126/science.372.6546.1021
39. Errors found in dozens of papers by top scientists at Dana-Farber Cancer Institute. 2024. doi:10.1126/science.z0oibzm
40. Bak-Coleman J, West J, O’Connor C, Bergstrom CT. Industry Influence in High-Profile Social Media Research. arXiv; 2026. doi:10.48550/ARXIV.2601.11507
41. Priem J, Piwowar H, Orr R. OpenAlex: A fully-open index of scholarly works, authors, venues, institutions, and concepts. arXiv; 2022. doi:10.48550/ARXIV.2205.01833
42. Ibrahim S, Zhang Y. AI-generated rat genitalia: Swiss publisher of scientific journal under pressure. swissinfo. 13 Mar 2024. Available: https://www.swissinfo.ch/eng/science/wrong-ai-generated-images-in-scientific-journal-put-a-strain-on-swiss-publisher-frontiers/73657004. Accessed 16 Apr 2026.
43. MDPI. Comment on: “Journal citation reports and the definition of a predatory journal: The case of the Multidisciplinary Digital Publishing Institute (MDPI)” from Oviedo-García. 2021 Aug. Report No.: 2979. Available: https://www.mdpi.com/about/announcements/2979

# The Issue with Special Issues: when Guest Editors Publish in Support of Self

## *Appendices & supplementary materials*

Below you will find al supplementary tables (S1-S7) and Figures (S1-S2). Two additional files complete the set of materials attached to this paper:

**File S1:** An example of a mass solicitation email sent to MAH, by MDPI Ecologies, where the carbon copy recipient list included all emails of all recipients, likely intended to be entered in the blind carbon copy field. Following a sarcastic response by MAH highlighting the preprint version of *The strain on scientific publishing* [1] as a possible submission, the MDPI Ecologies guest editor and MDPI staff assistant editor encouraged MAH to formally submit a manuscript to this special issue. It must be emphasised that MAH explained in his first reply, alongside providing a link to ref [1], that he has no relevant expertise to the topic at hand.

**File S2**: A collection of publisher-provided guidance on guest edited article collections that explicitly states that Frontiers and Elsevier will adjust the article metadata to remove the "special issue" classifier if the issue fails to reach a minimum of 4 articles, obscuring the conflict of interest and editorial context under which the article was assessed. In addition, webshots or PDF documents showing publisher policies that motivate guest editors to publish within the journal as a reward for guest editing, or within the special issue itself, are provided. The publisher's guest editor proposal forms are also provided for MDPI, Frontiers, and Phil Trans. Comments are provided for BMC and Discover regarding guest editor initial contact.

**Table S1. Prevalence of Special Issues across publishers, 2016-22.**

**Share of Special Issue papers out of total papers published**

Selected publishers · data from Hanson et al. (2024)

| | 2016 | 2017 | 2018 | 2019 | 2020 | 2021 | 2022 |
|---|---|---|---|---|---|---|---|
| **Overall: N papers** | 356K | 381K | 428K | 483K | 609K | 729K | 788K |
| **Overall: share SI** | 11% | 13% | 18% | 22% | 27% | 33% | 38% |
| **MDPI** | 74% | 75% | 78% | 81% | 86% | 88% | 88% |
| **Frontiers** | 49% | 45% | 52% | 60% | 61% | 66% | 69% |
| **Hindawi** | 25% | 24% | 27% | 20% | 28% | 51% | 62% |
| **BMC** | 9% | 9% | 10% | 10% | 12% | 12% | 10% |
| **Wiley** | 7% | 7% | 7% | 7% | 6% | 6% | 5% |
| **Nature** | 5% | 8% | 11% | 11% | 13% | 11% | 11% |
| **Springer** | 4% | 5% | 5% | 5% | 5% | 4% | 3% |
| **PLOS** | 4% | 5% | 4% | 5% | 9% | 11% | 1% |

# Table S2: Comparison of Special Issue (SI) Policies and Practices by Publisher

| | | | | | | | | Minimum N Papers in SI | | | Maximum allowed endogeny share | | |
|---|---|---|---|---|---|---|---|---|---|---|---|---|---|
| PUBLISHER | FOR-PROFIT | % SI PAPERS | NAME OF GUEST COLLECTIONS | METHOD | MINIMUM LENGTH OF INITIAL SUBMITTED PROPOSAL TEXT, INCLUDING PLANNED PAPER SUBMISSIONS (WORDS) † | PAPER RELEASE SCHEDULE | DOES PUBLISHER INVITE SUBMISSIONS INDEPENDENT OF THE GUEST EDITOR? | STATED POLICY | ACTUAL MIN | % NON-COMPLIANT | STATED POLICY | ACTUAL MAX | % NON-COMPLIANT |
| **MDPI** | Yes | 88%* | Special Issue | Web scraping | ~350 word proposal form | Rolling paper release as manuscripts are accepted | Yes, via email spam and calls for papers on website | 10 | 1 | 64% | None | 100% | — |
| **Frontiers** | Yes | 69%* | Research Topic | Web scraping | Only personal information requested at point of contact. | Rolling paper release as manuscripts are accepted | Yes, calls for papers are listed on website. The publisher will also suggest potential contributors to contact in the editorial centre for the editor to consider. | 4 | 1 | 2% | 30% | 100% | 8% |
| **Discover** | Yes | 42%* | Collection | Web scraping | Unknown | Rolling paper release as manuscripts are accepted | Yes, calls for papers are listed on website. Unknown if calls for papers are sent independent of the guest editor. | None | 1 | — | None | 100% | — |
| **BMC** | Yes | 10%* | Collection | Web scraping + manual curation | Special issues arise first from internal senior editorial staff. Guest editors are invited to contribute to editorial tasks, and to shape the final form of the issue. But guest editors do not necessarily propose topics. | Rolling paper release as manuscripts are accepted | Yes, calls for papers are listed on website | None | 1 | — | 25% | 100% | 16% |
| **Phil Trans** | No | 100% | Theme Issue | Web scraping with publisher cooperation | Pre-submission enquiry available to discuss. Initial forms of ~5500 words propose special issue contents. | Simultaneous upon final paper being ready | No | None | 6 | — | 3 papers/editor, 1 as lead author | 61% | — |

* SI papers as % of total papers from 2022, data from Hanson et al. (2024; QSS) and web scraping. Discover estimated from scraped data, with paper totals ~98% similar to totals from the Scopus database (Scimagojr.com).

† Initial submission to publisher, assuming one guest editor, estimated by filling in proposal forms with requested information.

**Figure S1: distribution of number of articles per special issue in our dataset, by publisher**

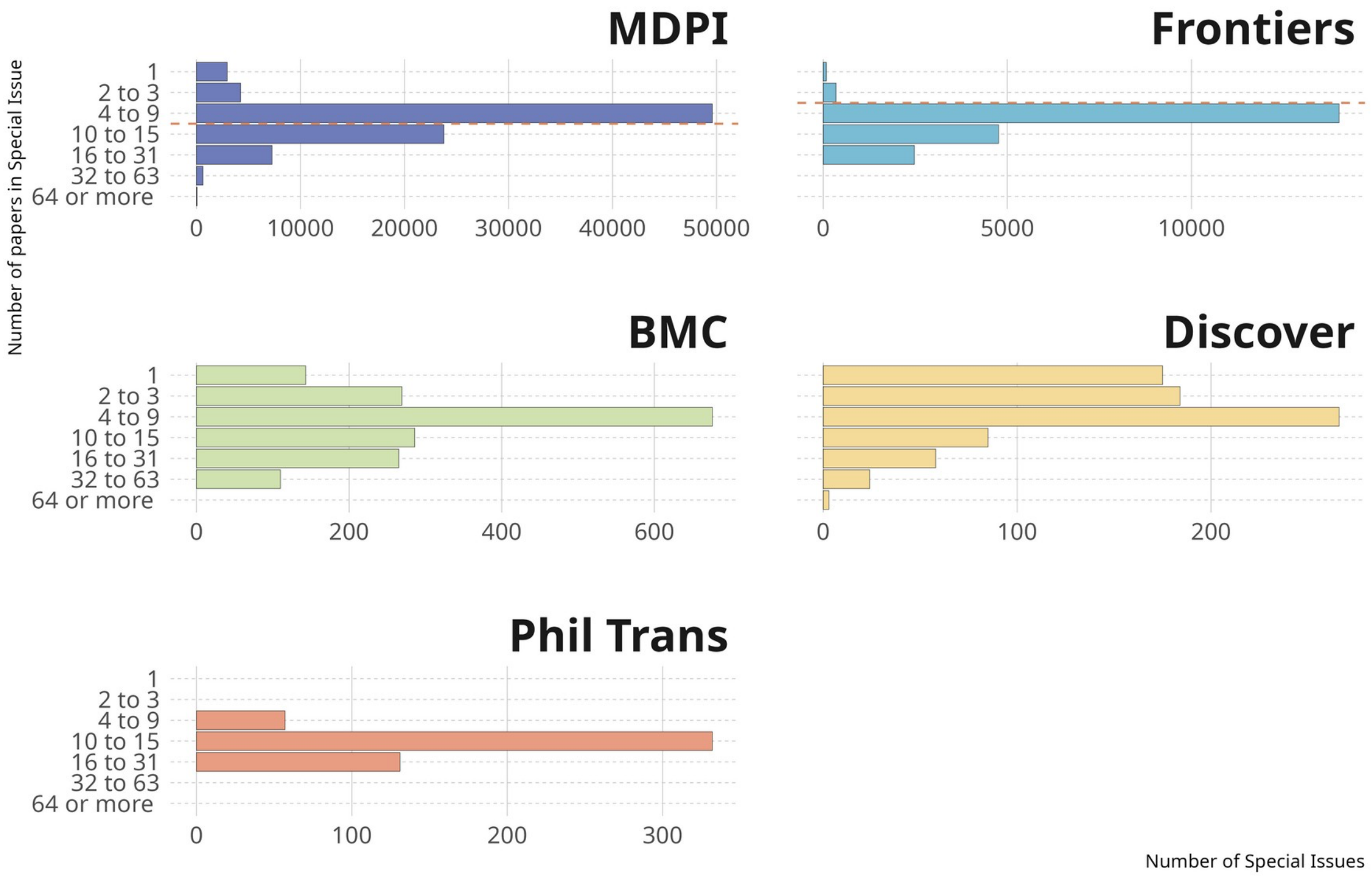

**Figure S2: mean endogeny levels over the years.** Grey lines indicate non-highlighted publishers to allow more direct comparisons.

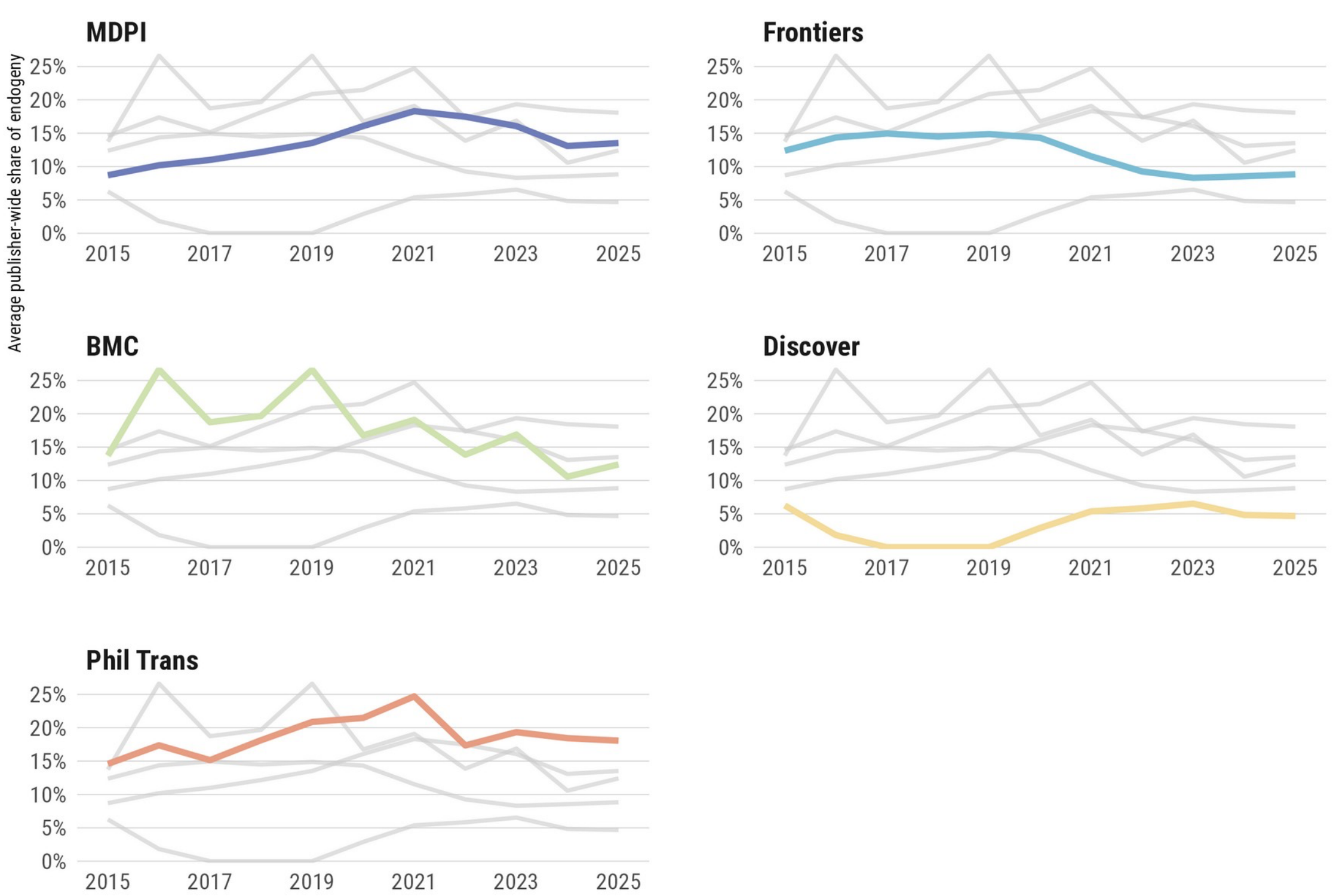

**Table S3: Individual journals endogeny and SI-hacking levels – Philosophical Transactions**

| | All SIs, 2015-25 | | | | 2024-25 only | | | |
|---|---|---|---|---|---|---|---|---|
| | N | Mean Endogeny | N hacked SI | SI-hacking rate | N | Mean Endogeny | N hacked SI | SI-hacking rate |
| PTB | 261 | 19.24% | 24 | 9.20% | 27 | 21.54% | 2 | 7.41% |
| PTA | 259 | 18.22% | 37 | 14.29% | 27 | 15.26% | 3 | 11.11% |

**Table S4: Individual journals endogeny and SI-hacking levels -- Discover**

| | All SIs, 2015-25 | | | | 2024-25 only | | | |
|---|---|---|---|---|---|---|---|---|
| | N | Mean Endogeny | N hacked SI | SI-hacking rate | N | Mean Endogeny | N hacked SI | SI-hacking rate |
| Discover Applied Sciences | 254 | 2.84% | 2 | 0.79% | 107 | 2.32% | 1 | 0.93% |
| Discover Sustainability | 131 | 3.54% | 3 | 2.29% | 122 | 3.67% | 3 | 2.46% |
| Discover Nano | 55 | 5.67% | 3 | 5.45% | 44 | 6.33% | 3 | 6.82% |
| Discover Water | 26 | 7.76% | 2 | 7.69% | 21 | 3.65% | 1 | 4.76% |
| Discover Internet of Things | 23 | 2.17% | 1 | 4.35% | 17 | 0.00% | 0 | 0.00% |
| Discover Food | 19 | 4.77% | 1 | 5.26% | 16 | 4.77% | 1 | 6.25% |
| Discover Materials | 14 | 4.18% | 1 | 7.14% | 7 | 5.49% | 1 | 14.29% |
| Discover Public Health | 14 | 15.71% | 2 | 14.29% | 14 | 15.71% | 2 | 14.29% |
| Discover Chemical Engineering | 13 | 7.95% | 1 | 7.69% | 9 | 6.67% | 1 | 11.11% |
| Discover Mechanical Engineering | 11 | 9.09% | 1 | 9.09% | 7 | 0.00% | 0 | 0.00% |
| Discover Environment | 9 | 11.11% | 1 | 11.11% | 9 | 11.11% | 1 | 11.11% |
| Discover Social Science and Health | 7 | 7.14% | 1 | 14.29% | 3 | 0.00% | 0 | 0.00% |
| Discover Analytics | 6 | 16.67% | 1 | 16.67% | 6 | 16.67% | 1 | 16.67% |
| Discover Medicine | 5 | 20.00% | 1 | 20.00% | 5 | 20.00% | 1 | 20.00% |
| Discover Chemistry | 4 | 25.00% | 1 | 25.00% | 4 | 25.00% | 1 | 25.00% |

**Table S5: Individual journals endogeny and SI-hacking levels – BMC**

| | All SIs, 2015-25 | | | | 2024-25 only | | | |
|---|---|---|---|---|---|---|---|---|
| | N | Mean Endogeny | N hacked SI | SI-hacking rate | N | Mean Endogeny | N hacked SI | SI-hacking rate |
| Inflammation and Regeneration | 42 | 15.25% | 2 | 4.76% | 2 | 10.00% | 0 | 0.00% |
| BMC Medicine | 39 | 9.76% | 1 | 2.56% | 9 | 3.42% | 0 | 0.00% |
| BMC Infectious Diseases | 29 | 3.37% | 1 | 3.45% | 19 | 1.84% | 0 | 0.00% |
| The Journal of Headache and Pain | 21 | 22.04% | 5 | 23.81% | 8 | 15.23% | 1 | 12.50% |
| BMC Pregnancy and Childbirth | 20 | 8.84% | 3 | 15.00% | 17 | 6.60% | 2 | 11.76% |
| Infectious Diseases of Poverty | 20 | 27.51% | 6 | 30.00% | 4 | 25.00% | 1 | 25.00% |
| International Journal for Equity in Health | 20 | 27.14% | 5 | 25.00% | 8 | 9.69% | 0 | 0.00% |
| Eye and Vision | 19 | 29.54% | 7 | 36.84% | 2 | 20.00% | 1 | 50.00% |
| BMC Medical Education | 17 | 8.10% | 1 | 5.88% | 16 | 7.82% | 1 | 6.25% |
| Journal of Biomedical Science | 17 | 11.29% | 1 | 5.88% | 1 | 0.00% | 0 | 0.00% |
| BMC Global and Public Health | 16 | 11.89% | 2 | 12.50% | 16 | 11.89% | 2 | 12.50% |
| Harm Reduction Journal | 16 | 10.56% | 2 | 12.50% | 5 | 18.57% | 1 | 20.00% |
| BMC Microbiology | 15 | 3.46% | 1 | 6.67% | 12 | 0.38% | 0 | 0.00% |
| BMC Chemistry | 14 | 19.05% | 3 | 21.43% | 10 | 16.67% | 2 | 20.00% |
| BMC Complementary Medicine and Therapies | 14 | 9.33% | 1 | 7.14% | 14 | 9.33% | 1 | 7.14% |
| BMC Palliative Care | 14 | 8.81% | 1 | 7.14% | 11 | 11.21% | 1 | 9.09% |
| Cell & Bioscience | 14 | 10.11% | 2 | 14.29% | 1 | 33.33% | 0 | 0.00% |
| Journal of Eating Disorders | 14 | 13.17% | 1 | 7.14% | 5 | 6.99% | 0 | 0.00% |
| BMC Ecology and Evolution | 13 | 6.21% | 1 | 7.69% | 10 | 8.08% | 1 | 10.00% |
| BMC Medical Informatics and Decision Making | 13 | 3.85% | 1 | 7.69% | 10 | 5.00% | 1 | 10.00% |
| BMC Nephrology | 13 | 9.52% | 1 | 7.69% | 12 | 8.33% | 1 | 8.33% |
| Energy, Sustainability and Society | 13 | 13.99% | 2 | 15.38% | 4 | 11.90% | 0 | 0.00% |
| Trials | 13 | 6.15% | 1 | 7.69% | 5 | 1.67% | 0 | 0.00% |
| Child and Adolescent Psychiatry and Mental Health | 12 | 15.97% | 2 | 16.67% | 2 | 0.00% | 0 | 0.00% |
| BMC Pulmonary Medicine | 11 | 11.04% | 2 | 18.18% | 9 | 5.56% | 1 | 11.11% |
| Journal of Animal Science and Biotechnology | 11 | 16.21% | 2 | 18.18% | 2 | 58.33% | 2 | 100.00% |
| Journal of Cheminformatics | 11 | 8.25% | 1 | 9.09% | 3 | 5.80% | 0 | 0.00% |
| BMC Biotechnology | 10 | 5.99% | 1 | 10.00% | 7 | 1.79% | 0 | 0.00% |
| CABI Agriculture and Bioscience | 10 | 9.00% | 2 | 20.00% | 4 | 10.00% | 1 | 25.00% |
| Stem Cell Research & Therapy | 10 | 5.25% | 1 | 10.00% | 4 | 10.00% | 1 | 25.00% |

| | All SIs, 2015-25 | | | | 2024-25 only | | | |
|---|---|---|---|---|---|---|---|---|
| | N | Mean Endogeny | N hacked SI | SI-hacking rate | N | Mean Endogeny | N hacked SI | SI-hacking rate |
| Arthroplasty | 9 | 14.63% | 1 | 11.11% | 3 | 6.67% | 0 | 0.00% |
| BMC Neuroscience | 9 | 17.08% | 2 | 22.22% | 8 | 18.75% | 2 | 25.00% |
| BioMedical Engineering OnLine | 9 | 16.24% | 2 | 22.22% | 2 | 29.76% | 1 | 50.00% |
| Immunity & Ageing | 9 | 18.06% | 1 | 11.11% | 3 | 6.67% | 0 | 0.00% |
| Reproductive Health | 9 | 8.15% | 1 | 11.11% | 2 | 0.00% | 0 | 0.00% |
| Alzheimer's Research & Therapy | 8 | 19.11% | 2 | 25.00% | 1 | 0.00% | 0 | 0.00% |
| BMC Medical Ethics | 8 | 8.65% | 1 | 12.50% | 5 | 0.00% | 0 | 0.00% |
| Journal of Neurodevelopmental Disorders | 8 | 10.92% | 1 | 12.50% | 1 | 0.00% | 0 | 0.00% |
| Journal of Venomous Animals and Toxins including Tropical Diseases | 8 | 15.15% | 1 | 12.50% | --- | --- | --- | --- |
| BMC Endocrine Disorders | 7 | 14.29% | 2 | 28.57% | 6 | 16.67% | 2 | 33.33% |
| BMC Genomic Data | 7 | 12.63% | 1 | 14.29% | 7 | 12.63% | 1 | 14.29% |
| BioPsychoSocial Medicine | 7 | 11.90% | 1 | 14.29% | --- | --- | --- | --- |
| Biomaterials Research | 7 | 20.58% | 1 | 14.29% | --- | --- | --- | --- |
| Burns & Trauma | 7 | 35.83% | 3 | 42.86% | --- | --- | --- | --- |
| Arthritis Research & Therapy | 6 | 8.18% | 1 | 16.67% | 2 | 0.00% | 0 | 0.00% |
| BMC Pharmacology and Toxicology | 6 | 6.67% | 1 | 16.67% | 5 | 0.00% | 0 | 0.00% |
| Implementation Science Communications | 6 | 22.92% | 1 | 16.67% | 3 | 4.17% | 0 | 0.00% |
| Journal of Biomedical Semantics | 6 | 12.50% | 1 | 16.67% | 1 | 0.00% | 0 | 0.00% |
| Biotechnology for Biofuels | 5 | 15.32% | 1 | 20.00% | --- | --- | --- | --- |
| Journal of Cotton Research | 5 | 29.27% | 1 | 20.00% | 3 | 31.62% | 1 | 33.33% |
| Journal of Physiological Anthropology | 5 | 20.50% | 2 | 40.00% | 1 | 40.00% | 1 | 100.00% |
| Thrombosis Journal | 5 | 24.00% | 1 | 20.00% | 2 | 50.00% | 1 | 50.00% |
| Annals of Forest Science | 4 | 25.00% | 2 | 50.00% | 2 | 0.00% | 0 | 0.00% |
| Chinese Journal of Cancer | 4 | 29.58% | 1 | 25.00% | --- | --- | --- | --- |
| Environmental Evidence | 4 | 15.00% | 1 | 25.00% | 1 | 50.00% | 1 | 100.00% |
| GigaScience | 4 | 32.50% | 1 | 25.00% | --- | --- | --- | --- |
| International Journal of Emergency Medicine | 4 | 28.14% | 2 | 50.00% | 4 | 28.14% | 2 | 50.00% |
| Journal of Translational Medicine | 4 | 14.17% | 1 | 25.00% | 3 | 18.89% | 1 | 33.33% |

| | All SIs, 2015-25 | | | | 2024-25 only | | | |
|---|---|---|---|---|---|---|---|---|
| | N | Mean Endogeny | N hacked SI | SI-hacking rate | N | Mean Endogeny | N hacked SI | SI-hacking rate |
| Cancer Nanotechnology | 3 | 29.76% | 1 | 33.33% | --- | --- | --- | --- |
| Molecular Horticulture | 3 | 21.67% | 1 | 33.33% | --- | --- | --- | --- |
| Orphanet Journal of Rare Diseases | 3 | 14.29% | 1 | 33.33% | 3 | 14.29% | 1 | 33.33% |
| Revista Chilena de Historia Natural | 3 | 21.43% | 1 | 33.33% | 1 | 0.00% | 0 | 0.00% |
| Veterinary Research | 3 | 26.41% | 2 | 66.67% | --- | --- | --- | --- |
| Annals of Occupational and Environmental Medicine | 2 | 68.75% | 2 | 100.00% | --- | --- | --- | --- |
| Avian Research | 2 | 58.33% | 2 | 100.00% | --- | --- | --- | --- |
| BMC International Health and Human Rights | 2 | 35.58% | 1 | 50.00% | --- | --- | --- | --- |
| Canadian Journal of Kidney Health and Disease | 2 | 50.00% | 1 | 50.00% | --- | --- | --- | --- |
| Disaster and Military Medicine | 2 | 75.00% | 2 | 100.00% | --- | --- | --- | --- |
| Environmental Health | 2 | 30.00% | 1 | 50.00% | 1 | 50.00% | 1 | 100.00% |
| Genes & Nutrition | 2 | 71.43% | 2 | 100.00% | 1 | 100.00% | 1 | 100.00% |
| Geochemical Transactions | 2 | 29.76% | 1 | 50.00% | 1 | 16.67% | 0 | 0.00% |
| Health Nanotechnology | 2 | 50.00% | 1 | 50.00% | 2 | 50.00% | 1 | 50.00% |
| Journal of Ecology and Environment | 2 | 37.50% | 1 | 50.00% | --- | --- | --- | --- |
| Journal of Ethnobiology and Ethnomedicine | 2 | 20.00% | 1 | 50.00% | 1 | 40.00% | 1 | 100.00% |
| Philosophy, Ethics, and Humanities in Medicine | 2 | 26.25% | 1 | 50.00% | 1 | 40.00% | 1 | 100.00% |
| Phytopathology Research | 2 | 50.00% | 1 | 50.00% | 1 | 100.00% | 1 | 100.00% |
| Women's Midlife Health | 2 | 20.00% | 1 | 50.00% | --- | --- | --- | --- |
| Advances in Rheumatology | 1 | 38.46% | 1 | 100.00% | 1 | 38.46% | 1 | 100.00% |
| Carbon Balance and Management | 1 | 35.71% | 1 | 100.00% | 1 | 35.71% | 1 | 100.00% |
| Echo Research & Practice | 1 | 66.67% | 1 | 100.00% | --- | --- | --- | --- |
| Emerging Themes in Epidemiology | 1 | 100.00% | 1 | 100.00% | --- | --- | --- | --- |
| International Journal of Mental Health Systems | 1 | 42.86% | 1 | 100.00% | --- | --- | --- | --- |
| Scoliosis and Spinal Disorders | 1 | 100.00% | 1 | 100.00% | --- | --- | --- | --- |

**Table S6: Individual journals endogeny and SI-hacking levels – Frontiers**

| | All SIs, 2015-25 | | | | 2024-25 only | | | |
|---|---|---|---|---|---|---|---|---|
| | N | Mean Endogeny | N hacked SI | SI-hacking rate | N | Mean Endogeny | N hacked SI | SI-hacking rate |
| Frontiers in Plant Science | 1386 | 9.32% | 73 | 5.27% | 179 | 7.59% | 7 | 3.91% |
| Frontiers in Immunology | 1345 | 9.74% | 65 | 4.83% | 214 | 8.27% | 14 | 6.54% |
| Frontiers in Microbiology | 1149 | 8.15% | 44 | 3.83% | 151 | 6.20% | 4 | 2.65% |
| Frontiers in Oncology | 1051 | 7.51% | 46 | 4.38% | 121 | 4.62% | 1 | 0.83% |
| Frontiers in Endocrinology | 912 | 8.41% | 51 | 5.59% | 172 | 5.97% | 6 | 3.49% |
| Frontiers in Pharmacology | 872 | 8.60% | 35 | 4.01% | 140 | 7.17% | 3 | 2.14% |
| Frontiers in Psychology | 834 | 9.10% | 38 | 4.56% | 125 | 8.87% | 4 | 3.20% |
| Frontiers in Medicine | 728 | 11.07% | 63 | 8.65% | 132 | 6.82% | 6 | 4.55% |
| Frontiers in Physiology | 578 | 11.34% | 36 | 6.23% | 52 | 10.14% | 3 | 5.77% |
| Frontiers in Psychiatry | 578 | 10.46% | 35 | 6.06% | 95 | 8.66% | 5 | 5.26% |
| Frontiers in Neurology | 534 | 11.33% | 41 | 7.68% | 71 | 11.25% | 9 | 12.68% |
| Frontiers in Cardiovascular Medicine | 522 | 8.58% | 31 | 5.94% | 87 | 7.82% | 4 | 4.60% |
| Frontiers in Cell and Developmental Biology | 501 | 8.27% | 19 | 3.79% | 66 | 8.26% | 4 | 6.06% |
| Frontiers in Nutrition | 482 | 7.07% | 18 | 3.73% | 91 | 4.47% | 1 | 1.10% |
| Frontiers in Bioengineering and Biotechnology | 471 | 9.42% | 21 | 4.46% | 73 | 5.84% | 0 | 0.00% |
| Frontiers in Chemistry | 468 | 11.39% | 36 | 7.69% | 71 | 11.47% | 9 | 12.68% |
| Frontiers in Public Health | 467 | 9.86% | 39 | 8.35% | 78 | 5.05% | 0 | 0.00% |
| Frontiers in Genetics | 465 | 5.66% | 19 | 4.09% | 75 | 4.96% | 2 | 2.67% |
| Frontiers in Marine Science | 462 | 16.87% | 65 | 14.07% | 55 | 14.12% | 4 | 7.27% |
| Frontiers in Veterinary Science | 424 | 12.08% | 39 | 9.20% | 76 | 7.88% | 3 | 3.95% |
| Frontiers in Cellular and Infection Microbiology | 394 | 6.80% | 10 | 2.54% | 106 | 5.55% | 3 | 2.83% |
| Frontiers in Energy Research | 368 | 10.26% | 23 | 6.25% | 62 | 5.64% | 1 | 1.61% |
| Frontiers in Neuroscience | 364 | 9.56% | 22 | 6.04% | 33 | 12.76% | 3 | 9.09% |
| Frontiers in Molecular Biosciences | 338 | 8.43% | 16 | 4.73% | 61 | 7.32% | 2 | 3.28% |
| Frontiers in Human Neuroscience | 301 | 11.46% | 22 | 7.31% | 45 | 12.93% | 3 | 6.67% |
| Frontiers in Ecology and Evolution | 275 | 14.02% | 29 | 10.55% | 21 | 8.64% | 1 | 4.76% |
| Frontiers in Earth Science | 270 | 13.61% | 27 | 10.00% | 31 | 13.01% | 4 | 12.90% |
| Frontiers in Environmental Science | 267 | 8.97% | 15 | 5.62% | 31 | 4.30% | 1 | 3.23% |
| Frontiers in Cellular Neuroscience | 255 | 8.25% | 5 | 1.96% | 41 | 6.17% | 1 | 2.44% |
| Frontiers in Pediatrics | 231 | 12.60% | 27 | 11.69% | 35 | 7.61% | 2 | 5.71% |

| | All SIs, 2015-25 | | | | 2024-25 only | | | |
|---|---|---|---|---|---|---|---|---|
| | N | Mean Endogeny | N hacked SI | SI-hacking rate | N | Mean Endogeny | N hacked SI | SI-hacking rate |
| Frontiers in Materials | 228 | 13.51% | 24 | 10.53% | 39 | 7.26% | 3 | 7.69% |
| Frontiers in Aging Neuroscience | 226 | 7.20% | 7 | 3.10% | 27 | 9.84% | 2 | 7.41% |
| Frontiers in Physics | 222 | 11.86% | 17 | 7.66% | 32 | 9.91% | 2 | 6.25% |
| Frontiers in Sustainable Food Systems | 209 | 12.35% | 17 | 8.13% | 48 | 10.89% | 1 | 2.08% |
| Frontiers in Molecular Neuroscience | 202 | 7.27% | 6 | 2.97% | 34 | 8.59% | 2 | 5.88% |
| Frontiers in Surgery | 172 | 11.15% | 16 | 9.30% | 36 | 10.15% | 4 | 11.11% |
| Frontiers in Behavioral Neuroscience | 149 | 9.33% | 8 | 5.37% | 7 | 9.29% | 1 | 14.29% |
| Frontiers in Education | 143 | 11.05% | 8 | 5.59% | 34 | 12.34% | 3 | 8.82% |
| Frontiers in Forests and Global Change | 137 | 15.00% | 17 | 12.41% | 18 | 14.60% | 3 | 16.67% |
| Frontiers in Sports and Active Living | 132 | 11.10% | 11 | 8.33% | 29 | 10.13% | 2 | 6.90% |
| Frontiers in Built Environment | 122 | 21.95% | 30 | 24.59% | 18 | 13.46% | 3 | 16.67% |
| Frontiers in Robotics and AI | 121 | 13.62% | 13 | 10.74% | 21 | 11.99% | 3 | 14.29% |
| Frontiers in Neurorobotics | 89 | 8.06% | 4 | 4.49% | 11 | 6.36% | 1 | 9.09% |
| Frontiers in Climate | 72 | 14.69% | 10 | 13.89% | 10 | 23.59% | 3 | 30.00% |
| Frontiers in Pain Research | 72 | 14.87% | 11 | 15.28% | 19 | 14.37% | 3 | 15.79% |
| Frontiers in Rehabilitation Sciences | 69 | 14.95% | 10 | 14.49% | 16 | 16.64% | 2 | 12.50% |
| Frontiers in Computer Science | 66 | 11.74% | 7 | 10.61% | 11 | 10.45% | 2 | 18.18% |
| Frontiers in Allergy | 62 | 12.24% | 5 | 8.06% | 22 | 10.34% | 1 | 4.55% |
| Frontiers in Communication | 58 | 8.00% | 1 | 1.72% | 14 | 5.94% | 0 | 0.00% |
| Frontiers in Political Science | 56 | 14.00% | 3 | 5.36% | 17 | 9.64% | 0 | 0.00% |
| Frontiers in Virtual Reality | 56 | 9.55% | 4 | 7.14% | 13 | 8.97% | 1 | 7.69% |
| Frontiers in Computational Neuroscience | 55 | 6.43% | 2 | 3.64% | 6 | 0.00% | 0 | 0.00% |
| Frontiers in Sociology | 55 | 13.13% | 7 | 12.73% | 21 | 9.09% | 3 | 14.29% |
| Frontiers in Digital Health | 53 | 12.47% | 5 | 9.43% | 12 | 15.42% | 1 | 8.33% |
| Frontiers in Systems Neuroscience | 47 | 6.83% | 2 | 4.26% | 3 | 6.67% | 0 | 0.00% |
| Frontiers in Applied Mathematics and Statistics | 43 | 9.86% | 3 | 6.98% | 6 | 13.65% | 0 | 0.00% |
| Frontiers in Mechanical Engineering | 42 | 20.01% | 10 | 23.81% | 5 | 19.22% | 1 | 20.00% |
| Frontiers in Bioinformatics | 41 | 7.89% | 3 | 7.32% | 10 | 16.33% | 2 | 20.00% |

| | All SIs, 2015-25 | | | | 2024-25 only | | | |
|---|---|---|---|---|---|---|---|---|
| | N | Mean Endogeny | N hacked SI | SI-hacking rate | N | Mean Endogeny | N hacked SI | SI-hacking rate |
| Frontiers in Neuroanatomy | 41 | 13.98% | 4 | 9.76% | 3 | 21.48% | 0 | 0.00% |
| Frontiers in Toxicology | 41 | 16.19% | 7 | 17.07% | 13 | 9.54% | 1 | 7.69% |
| Frontiers in Research Metrics and Analytics | 40 | 10.30% | 2 | 5.00% | 8 | 13.75% | 1 | 12.50% |
| Frontiers in Sustainability | 40 | 18.84% | 6 | 15.00% | 5 | 21.19% | 1 | 20.00% |
| Frontiers in Aging | 39 | 6.99% | 2 | 5.13% | 11 | 7.02% | 1 | 9.09% |
| Frontiers in Artificial Intelligence | 39 | 11.01% | 4 | 10.26% | 8 | 10.95% | 0 | 0.00% |
| Frontiers in Reproductive Health | 39 | 13.20% | 3 | 7.69% | 7 | 22.89% | 1 | 14.29% |
| Frontiers in Fungal Biology | 38 | 10.73% | 5 | 13.16% | 1 | 0.00% | 0 | 0.00% |
| Frontiers in Global Women's Health | 37 | 7.35% | 1 | 2.70% | 7 | 4.64% | 0 | 0.00% |
| Frontiers in Animal Science | 32 | 11.35% | 2 | 6.25% | 5 | 5.43% | 0 | 0.00% |
| Frontiers in Dental Medicine | 32 | 15.12% | 5 | 15.62% | 2 | 16.67% | 0 | 0.00% |
| Frontiers in Chemical Engineering | 29 | 16.38% | 3 | 10.34% | 4 | 11.44% | 0 | 0.00% |
| Frontiers in Synaptic Neuroscience | 29 | 8.70% | 1 | 3.45% | 3 | 21.67% | 1 | 33.33% |
| Frontiers in Big Data | 27 | 10.70% | 2 | 7.41% | 4 | 0.00% | 0 | 0.00% |
| Frontiers in Medical Technology | 26 | 9.90% | 3 | 11.54% | 5 | 14.67% | 1 | 20.00% |
| Frontiers in Network Physiology | 26 | 18.27% | 3 | 11.54% | 7 | 25.00% | 1 | 14.29% |
| Frontiers in Ophthalmology | 23 | 8.54% | 2 | 8.70% | 9 | 10.12% | 1 | 11.11% |
| Frontiers in Blockchain | 22 | 8.51% | 1 | 4.55% | 2 | 12.50% | 0 | 0.00% |
| Frontiers in Nephrology | 21 | 13.26% | 2 | 9.52% | 9 | 7.80% | 1 | 11.11% |
| Frontiers in Agronomy | 19 | 11.57% | 1 | 5.26% | 7 | 8.63% | 0 | 0.00% |
| Frontiers in Genome Editing | 18 | 6.86% | 1 | 5.56% | 3 | 8.33% | 0 | 0.00% |
| Frontiers in Tropical Diseases | 18 | 9.80% | 1 | 5.56% | 16 | 11.03% | 1 | 6.25% |
| Frontiers in Urology | 18 | 22.31% | 4 | 22.22% | 3 | 33.33% | 1 | 33.33% |
| Frontiers in Clinical Diabetes and Healthcare | 17 | 9.14% | 2 | 11.76% | 10 | 6.50% | 1 | 10.00% |
| Frontiers in Epidemiology | 17 | 13.40% | 2 | 11.76% | 3 | 3.70% | 0 | 0.00% |
| Frontiers in Remote Sensing | 17 | 23.02% | 6 | 35.29% | 3 | 34.21% | 2 | 66.67% |
| Frontiers in Neuroergonomics | 16 | 8.81% | 2 | 12.50% | 3 | 0.00% | 0 | 0.00% |
| Frontiers in Virology | 15 | 5.67% | 1 | 6.67% | 6 | 4.17% | 0 | 0.00% |
| Frontiers in Astronomy and Space Sciences | 14 | 16.14% | 2 | 14.29% | 1 | 50.00% | 1 | 100.00% |

| | All SIs, 2015-25 | | | | 2024-25 only | | | |
|---|---|---|---|---|---|---|---|---|
| | N | Mean Endogeny | N hacked SI | SI-hacking rate | N | Mean Endogeny | N hacked SI | SI-hacking rate |
| Frontiers in Insect Science | 14 | 28.83% | 3 | 21.43% | 5 | 32.00% | 1 | 20.00% |
| Frontiers in Transplantation | 14 | 13.21% | 3 | 21.43% | 8 | 23.12% | 3 | 37.50% |
| Frontiers in Radiology | 13 | 17.89% | 2 | 15.38% | 3 | 20.45% | 1 | 33.33% |
| Frontiers in Signal Processing | 13 | 16.54% | 3 | 23.08% | 4 | 10.00% | 0 | 0.00% |
| Frontiers in Dementia | 11 | 14.39% | 2 | 18.18% | 8 | 12.50% | 2 | 25.00% |
| Frontiers in Child and Adolescent Psychiatry | 10 | 10.74% | 1 | 10.00% | 5 | 10.33% | 0 | 0.00% |
| Frontiers in Hematology | 10 | 17.43% | 1 | 10.00% | 9 | 16.59% | 1 | 11.11% |
| Frontiers in Molecular Medicine | 10 | 16.61% | 1 | 10.00% | 3 | 15.00% | 0 | 0.00% |
| Frontiers in Nuclear Medicine | 10 | 15.00% | 3 | 30.00% | 3 | 13.33% | 1 | 33.33% |
| Frontiers in Analytical Science | 8 | 14.68% | 1 | 12.50% | 1 | 9.09% | 0 | 0.00% |
| Frontiers in Future Transportation | 8 | 22.92% | 1 | 12.50% | --- | --- | --- | --- |
| Frontiers in Control Engineering | 7 | 23.88% | 2 | 28.57% | 1 | 25.00% | 0 | 0.00% |
| Frontiers in Soft Matter | 7 | 21.19% | 2 | 28.57% | 3 | 8.33% | 0 | 0.00% |
| Frontiers in Stroke | 7 | 21.56% | 1 | 14.29% | 5 | 23.33% | 1 | 20.00% |
| Frontiers in Sustainable Tourism | 7 | 25.48% | 1 | 14.29% | 4 | 34.17% | 1 | 25.00% |
| Frontiers in Environmental Economics | 6 | 20.71% | 1 | 16.67% | 1 | 14.29% | 0 | 0.00% |
| Frontiers in Food Science and Technology | 6 | 15.83% | 1 | 16.67% | 4 | 5.00% | 0 | 0.00% |
| Frontiers in Space Technologies | 6 | 15.83% | 1 | 16.67% | 1 | 50.00% | 1 | 100.00% |
| Frontiers in Drug Delivery | 5 | 10.00% | 1 | 20.00% | 1 | 0.00% | 0 | 0.00% |
| Frontiers in Drug Safety and Regulation | 5 | 17.83% | 1 | 20.00% | 2 | 10.00% | 0 | 0.00% |
| Frontiers in Electronics | 5 | 26.67% | 2 | 40.00% | --- | --- | --- | --- |
| Frontiers in Environmental Chemistry | 5 | 21.64% | 1 | 20.00% | 2 | 12.50% | 0 | 0.00% |
| Frontiers in Bee Science | 4 | 21.07% | 1 | 25.00% | 4 | 21.07% | 1 | 25.00% |
| Frontiers in Horticulture | 4 | 40.62% | 2 | 50.00% | 3 | 45.83% | 2 | 66.67% |
| Frontiers in Manufacturing Technology | 4 | 35.42% | 2 | 50.00% | --- | --- | --- | --- |
| Frontiers in Natural Products | 4 | 36.25% | 2 | 50.00% | 2 | 10.00% | 0 | 0.00% |
| Frontiers in Aquaculture | 3 | 13.89% | 1 | 33.33% | 2 | 20.83% | 1 | 50.00% |
| Frontiers in Complex Systems | 3 | 13.33% | 1 | 33.33% | 3 | 13.33% | 1 | 33.33% |

| | All SIs, 2015-25 | | | | 2024-25 only | | | |
|---|---|---|---|---|---|---|---|---|
| | N | Mean Endogeny | N hacked SI | SI-hacking rate | N | Mean Endogeny | N hacked SI | SI-hacking rate |
| Frontiers in Developmental Psychology | 3 | 31.85% | 1 | 33.33% | 3 | 31.85% | 1 | 33.33% |
| Frontiers in Digital Humanities | 3 | 14.81% | 1 | 33.33% | --- | --- | --- | --- |
| Frontiers in Ethology | 3 | 23.33% | 1 | 33.33% | 3 | 23.33% | 1 | 33.33% |
| Frontiers in Audiology and Otology | 2 | 38.64% | 1 | 50.00% | 2 | 38.64% | 1 | 50.00% |
| Frontiers in Communications and Networks | 1 | 75.00% | 1 | 100.00% | --- | --- | --- | --- |

**Table S7: Individual journals endogeny and SI-hacking levels – MDPI**

| | All SIs, 2015-25 | | | | 2024-25 only | | | |
|---|---|---|---|---|---|---|---|---|
| | N | Mean Endogeny | N hacked SI | SI-hacking rate | N | Mean Endogeny | N hacked SI | SI-hacking rate |
| International Journal of Molecular Sciences | 6734 | 12.59% | 478 | 7.10% | 1954 | 12.49% | 140 | 7.16% |
| Sustainability | 4412 | 16.61% | 567 | 12.85% | 824 | 12.49% | 49 | 5.95% |
| Applied Sciences | 3976 | 15.53% | 441 | 11.09% | 1135 | 10.79% | 53 | 4.67% |
| Molecules | 3066 | 12.37% | 185 | 6.03% | 690 | 9.74% | 21 | 3.04% |
| Energies | 2915 | 19.57% | 485 | 16.64% | 634 | 15.31% | 49 | 7.73% |
| International Journal of Environmental Research and Public Health | 2913 | 17.89% | 366 | 12.56% | 199 | 14.78% | 11 | 5.53% |
| Sensors | 2631 | 11.42% | 182 | 6.92% | 743 | 9.32% | 29 | 3.90% |
| Materials | 2539 | 18.57% | 378 | 14.89% | 722 | 14.83% | 72 | 9.97% |
| Journal of Clinical Medicine | 2179 | 17.86% | 228 | 10.46% | 848 | 15.17% | 49 | 5.78% |
| Remote Sensing | 2058 | 16.23% | 275 | 13.36% | 499 | 13.25% | 40 | 8.02% |
| Cancers | 1888 | 14.25% | 190 | 10.06% | 507 | 14.17% | 51 | 10.06% |
| Nutrients | 1850 | 16.74% | 217 | 11.73% | 571 | 15.74% | 51 | 8.93% |
| Polymers | 1703 | 14.64% | 135 | 7.93% | 506 | 13.76% | 41 | 8.10% |
| Water | 1512 | 21.80% | 309 | 20.44% | 365 | 17.76% | 44 | 12.05% |
| Electronics | 1382 | 14.21% | 139 | 10.06% | 570 | 10.47% | 33 | 5.79% |
| Mathematics | 1360 | 14.01% | 97 | 7.13% | 484 | 12.96% | 33 | 6.82% |
| Foods | 1327 | 19.73% | 177 | 13.34% | 494 | 20.34% | 67 | 13.56% |
| Nanomaterials | 1268 | 16.63% | 135 | 10.65% | 268 | 18.82% | 36 | 13.43% |
| Plants | 1188 | 17.83% | 153 | 12.88% | 441 | 16.96% | 46 | 10.43% |
| Microorganisms | 1107 | 13.86% | 112 | 10.12% | 355 | 10.86% | 22 | 6.20% |
| Cells | 1078 | 13.58% | 85 | 7.88% | 215 | 13.32% | 19 | 8.84% |
| Biomedicines | 1034 | 12.06% | 74 | 7.16% | 438 | 10.87% | 32 | 7.31% |
| Animals | 1007 | 15.40% | 93 | 9.24% | 317 | 12.14% | 13 | 4.10% |
| Diagnostics | 990 | 15.31% | 80 | 8.08% | 403 | 13.35% | 29 | 7.20% |
| Viruses | 959 | 11.53% | 58 | 6.05% | 256 | 8.95% | 9 | 3.52% |
| Pharmaceutics | 898 | 14.04% | 68 | 7.57% | 240 | 14.69% | 19 | 7.92% |
| Agronomy | 851 | 17.70% | 118 | 13.87% | 251 | 14.13% | 18 | 7.17% |
| Entropy | 786 | 12.40% | 50 | 6.36% | 163 | 11.73% | 10 | 6.13% |
| Metals | 784 | 20.45% | 126 | 16.07% | 198 | 18.55% | 21 | 10.61% |
| Forests | 780 | 16.42% | 90 | 11.54% | 235 | 15.07% | 18 | 7.66% |
| Symmetry | 772 | 13.79% | 68 | 8.81% | 155 | 12.98% | 14 | 9.03% |
| Genes | 762 | 16.20% | 82 | 10.76% | 180 | 15.87% | 17 | 9.44% |
| Micromachines | 762 | 13.75% | 59 | 7.74% | 178 | 11.72% | 7 | 3.93% |
| Processes | 749 | 14.74% | 58 | 7.74% | 281 | 15.78% | 31 | 11.03% |
| Religions | 747 | 8.15% | 13 | 1.74% | 249 | 7.69% | 4 | 1.61% |

| | All SIs, 2015-25 | | | | 2024-25 only | | | |
|---|---|---|---|---|---|---|---|---|
| | N | Mean Endogeny | N hacked SI | SI-hacking rate | N | Mean Endogeny | N hacked SI | SI-hacking rate |
| Catalysts | 731 | 17.80% | 93 | 12.72% | 153 | 19.65% | 28 | 18.30% |
| Antioxidants | 715 | 12.17% | 42 | 5.87% | 178 | 9.89% | 8 | 4.49% |
| Atmosphere | 714 | 14.49% | 77 | 10.78% | 184 | 12.64% | 12 | 6.52% |
| Biomolecules | 645 | 15.12% | 67 | 10.39% | 165 | 13.36% | 13 | 7.88% |
| Crystals | 636 | 16.73% | 81 | 12.74% | 144 | 15.17% | 14 | 9.72% |
| Minerals | 633 | 21.66% | 127 | 20.06% | 141 | 18.94% | 16 | 11.35% |
| Journal of Marine Science and Engineering | 622 | 21.92% | 133 | 21.38% | 227 | 18.67% | 32 | 14.10% |
| Coatings | 610 | 17.33% | 71 | 11.64% | 200 | 16.34% | 20 | 10.00% |
| Antibiotics | 608 | 15.61% | 65 | 10.69% | 183 | 12.33% | 9 | 4.92% |
| Buildings | 595 | 19.76% | 74 | 12.44% | 336 | 21.02% | 46 | 13.69% |
| Land | 583 | 12.25% | 46 | 7.89% | 235 | 8.29% | 4 | 1.70% |
| Life | 569 | 14.25% | 51 | 8.96% | 209 | 12.69% | 12 | 5.74% |
| Pharmaceuticals | 546 | 12.04% | 33 | 6.04% | 227 | 11.84% | 14 | 6.17% |
| Healthcare | 540 | 14.96% | 50 | 9.26% | 211 | 14.11% | 11 | 5.21% |
| Brain Sciences | 513 | 18.53% | 76 | 14.81% | 153 | 17.45% | 15 | 9.80% |
| Pathogens | 513 | 16.59% | 62 | 12.09% | 142 | 13.25% | 8 | 5.63% |
| Agriculture | 506 | 18.08% | 68 | 13.44% | 173 | 15.75% | 20 | 11.56% |
| Vaccines | 451 | 12.07% | 34 | 7.54% | 164 | 9.46% | 8 | 4.88% |
| Education Sciences | 440 | 11.40% | 25 | 5.68% | 206 | 8.74% | 6 | 2.91% |
| Children | 430 | 13.45% | 22 | 5.12% | 196 | 13.24% | 12 | 6.12% |
| Medicina | 423 | 21.80% | 83 | 19.62% | 159 | 19.05% | 17 | 10.69% |
| Toxins | 400 | 20.43% | 75 | 18.75% | 62 | 21.14% | 13 | 20.97% |
| Journal of Fungi | 392 | 14.57% | 35 | 8.93% | 122 | 12.85% | 8 | 6.56% |
| Marine Drugs | 388 | 13.41% | 24 | 6.19% | 91 | 10.80% | 2 | 2.20% |
| Metabolites | 378 | 17.00% | 48 | 12.70% | 96 | 17.50% | 15 | 15.62% |
| Journal of Personalized Medicine | 374 | 19.90% | 52 | 13.90% | 105 | 21.54% | 21 | 20.00% |
| Diversity | 356 | 17.37% | 35 | 9.83% | 98 | 14.85% | 4 | 4.08% |
| Bioengineering | 350 | 14.81% | 31 | 8.86% | 182 | 15.22% | 14 | 7.69% |
| Membranes | 331 | 18.35% | 41 | 12.39% | 37 | 25.14% | 6 | 16.22% |
| Photonics | 325 | 13.73% | 32 | 9.85% | 136 | 13.28% | 15 | 11.03% |
| Biosensors | 321 | 12.59% | 22 | 6.85% | 99 | 10.34% | 5 | 5.05% |
| Biology | 314 | 17.92% | 39 | 12.42% | 104 | 16.15% | 9 | 8.65% |
| Gels | 292 | 8.83% | 8 | 2.74% | 155 | 8.40% | 4 | 2.58% |
| Fermentation | 284 | 9.92% | 7 | 2.46% | 113 | 10.34% | 3 | 2.65% |
| Horticulturae | 284 | 12.94% | 20 | 7.04% | 146 | 11.59% | 6 | 4.11% |
| Axioms | 256 | 10.45% | 7 | 2.73% | 94 | 8.20% | 1 | 1.06% |

| | All SIs, 2015-25 | | | | 2024-25 only | | | |
|---|---|---|---|---|---|---|---|---|
| | N | Mean Endogeny | N hacked SI | SI-hacking rate | N | Mean Endogeny | N hacked SI | SI-hacking rate |
| Information | 247 | 11.20% | 17 | 6.88% | 96 | 9.65% | 4 | 4.17% |
| Insects | 246 | 15.79% | 28 | 11.38% | 56 | 8.96% | 1 | 1.79% |
| Aerospace | 240 | 15.39% | 26 | 10.83% | 96 | 13.74% | 4 | 4.17% |
| Toxics | 235 | 20.06% | 43 | 18.30% | 93 | 19.18% | 14 | 15.05% |
| Machines | 224 | 17.28% | 27 | 12.05% | 93 | 16.57% | 11 | 11.83% |
| Journal of Risk and Financial Management | 217 | 12.87% | 16 | 7.37% | 63 | 9.09% | 2 | 3.17% |
| Geosciences | 215 | 18.03% | 42 | 19.53% | 29 | 13.34% | 5 | 17.24% |
| Social Sciences | 212 | 10.41% | 12 | 5.66% | 65 | 11.00% | 4 | 6.15% |
| ISPRS International Journal of Geo-Information | 208 | 11.40% | 15 | 7.21% | 8 | 3.70% | 0 | 0.00% |
| Algorithms | 207 | 9.25% | 14 | 6.76% | 69 | 7.84% | 1 | 1.45% |
| Universe | 196 | 10.69% | 13 | 6.63% | 55 | 6.71% | 0 | 0.00% |
| Behavioral Sciences | 192 | 17.06% | 23 | 11.98% | 112 | 13.18% | 8 | 7.14% |
| Biomimetics | 192 | 10.58% | 12 | 6.25% | 109 | 11.18% | 8 | 7.34% |
| Fractal and Fractional | 188 | 12.03% | 8 | 4.26% | 79 | 9.82% | 2 | 2.53% |
| Chemosensors | 173 | 10.79% | 7 | 4.05% | 57 | 11.00% | 3 | 5.26% |
| Veterinary Sciences | 167 | 16.86% | 18 | 10.78% | 76 | 14.18% | 5 | 6.58% |
| Lubricants | 161 | 14.98% | 13 | 8.07% | 53 | 17.43% | 4 | 7.55% |
| Current Issues in Molecular Biology | 154 | 5.51% | 4 | 2.60% | 107 | 6.68% | 4 | 3.74% |
| Languages | 153 | 8.13% | 1 | 0.65% | 50 | 9.35% | 1 | 2.00% |
| Fluids | 150 | 11.39% | 5 | 3.33% | 39 | 7.66% | 0 | 0.00% |
| Journal of Functional Biomaterials | 141 | 18.27% | 16 | 11.35% | 77 | 16.14% | 7 | 9.09% |
| Future Internet | 140 | 10.63% | 12 | 8.57% | 43 | 6.37% | 1 | 2.33% |
| Inorganics | 140 | 10.86% | 2 | 1.43% | 43 | 8.80% | 0 | 0.00% |
| Tropical Medicine and Infectious Disease | 136 | 12.80% | 14 | 10.29% | 37 | 8.92% | 1 | 2.70% |
| Actuators | 135 | 16.99% | 24 | 17.78% | 48 | 12.49% | 4 | 8.33% |
| Humanities | 130 | 5.69% | 2 | 1.54% | 28 | 4.04% | 0 | 0.00% |
| Drones | 116 | 10.41% | 2 | 1.72% | 67 | 10.28% | 2 | 2.99% |
| Fishes | 114 | 19.78% | 18 | 15.79% | 62 | 21.06% | 11 | 17.74% |
| Heritage | 112 | 14.80% | 13 | 11.61% | 41 | 14.35% | 4 | 9.76% |
| Arts | 110 | 6.45% | 1 | 0.91% | 25 | 5.70% | 1 | 4.00% |
| Batteries | 108 | 18.43% | 9 | 8.33% | 48 | 16.03% | 3 | 6.25% |
| Separations | 108 | 17.02% | 10 | 9.26% | 40 | 21.33% | 6 | 15.00% |
| Journal of Cardiovascular Development and Disease | 107 | 14.13% | 9 | 8.41% | 49 | 11.53% | 1 | 2.04% |

| | All SIs, 2015-25 | | | | 2024-25 only | | | |
|---|---|---|---|---|---|---|---|---|
| | N | Mean Endogeny | N hacked SI | SI-hacking rate | N | Mean Endogeny | N hacked SI | SI-hacking rate |
| Journal of Imaging | 106 | 11.63% | 6 | 5.66% | 33 | 11.12% | 1 | 3.03% |
| Journal of Composites Science | 100 | 11.39% | 11 | 11.00% | 40 | 5.58% | 1 | 2.50% |
| Administrative Sciences | 96 | 15.45% | 10 | 10.42% | 43 | 11.90% | 2 | 4.65% |
| Systems | 96 | 12.04% | 6 | 6.25% | 50 | 8.93% | 1 | 2.00% |
| Societies | 93 | 13.04% | 3 | 3.23% | 27 | 12.00% | 0 | 0.00% |
| Environments | 92 | 11.99% | 9 | 9.78% | 36 | 4.29% | 0 | 0.00% |
| Current Oncology | 91 | 15.37% | 10 | 10.99% | 34 | 10.76% | 2 | 5.88% |
| Fire | 90 | 12.91% | 8 | 8.89% | 57 | 10.22% | 2 | 3.51% |
| Risks | 88 | 11.73% | 9 | 10.23% | 15 | 5.73% | 0 | 0.00% |
| Computers | 86 | 6.66% | 3 | 3.49% | 32 | 4.09% | 0 | 0.00% |
| Sports | 85 | 18.62% | 12 | 14.12% | 27 | 9.63% | 0 | 0.00% |
| Dentistry Journal | 78 | 16.52% | 10 | 12.82% | 32 | 11.84% | 1 | 3.12% |
| Climate | 76 | 13.40% | 6 | 7.89% | 13 | 14.65% | 1 | 7.69% |
| Philosophies | 75 | 9.57% | 2 | 2.67% | 26 | 6.41% | 0 | 0.00% |
| Computation | 72 | 15.22% | 10 | 13.89% | 25 | 11.77% | 2 | 8.00% |
| Pharmacy | 71 | 14.17% | 8 | 11.27% | 12 | 13.36% | 1 | 8.33% |
| Genealogy | 70 | 8.73% | 2 | 2.86% | 24 | 12.17% | 2 | 8.33% |
| Magnetochemistry | 69 | 17.49% | 8 | 11.59% | 14 | 13.77% | 1 | 7.14% |
| Hydrology | 65 | 11.30% | 3 | 4.62% | 22 | 7.55% | 1 | 4.55% |
| Infrastructures | 65 | 17.34% | 12 | 18.46% | 19 | 8.40% | 1 | 5.26% |
| Journal of Functional Morphology and Kinesiology | 64 | 11.01% | 6 | 9.38% | 29 | 7.95% | 0 | 0.00% |
| Journal of Manufacturing and Materials Processing | 62 | 10.35% | 4 | 6.45% | 23 | 5.16% | 0 | 0.00% |
| Resources | 62 | 13.57% | 6 | 9.68% | 12 | 11.05% | 0 | 0.00% |
| Atoms | 57 | 17.96% | 9 | 15.79% | 9 | 9.41% | 0 | 0.00% |
| Cosmetics | 57 | 18.23% | 8 | 14.04% | 18 | 15.91% | 2 | 11.11% |
| World Electric Vehicle Journal | 57 | 10.96% | 3 | 5.26% | 27 | 7.17% | 0 | 0.00% |
| Galaxies | 56 | 14.57% | 6 | 10.71% | 9 | 14.86% | 0 | 0.00% |
| Fibers | 53 | 14.82% | 6 | 11.32% | 5 | 6.86% | 0 | 0.00% |
| Robotics | 53 | 10.43% | 1 | 1.89% | 17 | 6.12% | 0 | 0.00% |
| Inventions | 52 | 10.30% | 4 | 7.69% | 7 | 5.24% | 0 | 0.00% |
| Technologies | 52 | 11.05% | 4 | 7.69% | 13 | 5.78% | 0 | 0.00% |
| Laws | 50 | 7.84% | 3 | 6.00% | 10 | 3.94% | 0 | 0.00% |
| Beverages | 48 | 11.66% | 3 | 6.25% | 5 | 28.34% | 1 | 20.00% |

| | All SIs, 2015-25 | | | | 2024-25 only | | | |
|---|---|---|---|---|---|---|---|---|
| | N | Mean Endogeny | N hacked SI | SI-hacking rate | N | Mean Endogeny | N hacked SI | SI-hacking rate |
| Big Data and Cognitive Computing | 41 | 12.07% | 5 | 12.20% | 18 | 6.51% | 1 | 5.56% |
| C | 40 | 10.17% | 2 | 5.00% | 11 | 4.96% | 0 | 0.00% |
| Diseases | 40 | 14.58% | 5 | 12.50% | 10 | 12.74% | 1 | 10.00% |
| Data | 39 | 13.12% | 5 | 12.82% | 5 | 20.00% | 2 | 40.00% |
| Games | 39 | 8.26% | 2 | 5.13% | 1 | 0.00% | 0 | 0.00% |
| Condensed Matter | 38 | 21.19% | 8 | 21.05% | 5 | 18.94% | 1 | 20.00% |
| Multimodal Technologies and Interaction | 38 | 8.16% | 1 | 2.63% | 11 | 2.94% | 0 | 0.00% |
| Mathematical and Computational Applications | 36 | 15.84% | 2 | 5.56% | 6 | 6.28% | 0 | 0.00% |
| Geriatrics | 35 | 14.13% | 1 | 2.86% | 5 | 2.86% | 0 | 0.00% |
| Medicines | 35 | 15.17% | 6 | 17.14% | --- | --- | --- | --- |
| International Journal of Financial Studies | 33 | 8.42% | 2 | 6.06% | 6 | 10.72% | 1 | 16.67% |
| International Journal of Neonatal Screening | 33 | 17.84% | 5 | 15.15% | 6 | 8.73% | 0 | 0.00% |
| Colloids and Interfaces | 32 | 15.24% | 2 | 6.25% | 9 | 5.80% | 0 | 0.00% |
| European Journal of Investigation in Health, Psychology and Education | 32 | 13.11% | 5 | 15.62% | 6 | 8.33% | 0 | 0.00% |
| Soil Systems | 32 | 13.64% | 3 | 9.38% | 8 | 4.23% | 0 | 0.00% |
| Urban Science | 32 | 11.68% | 2 | 6.25% | 7 | 0.00% | 0 | 0.00% |
| ChemEngineering | 31 | 16.60% | 6 | 19.35% | 6 | 12.23% | 1 | 16.67% |
| Antibodies | 30 | 13.77% | 3 | 10.00% | 4 | 8.35% | 0 | 0.00% |
| Ceramics | 29 | 9.74% | 2 | 6.90% | 8 | 5.99% | 0 | 0.00% |
| Designs | 29 | 8.46% | 1 | 3.45% | 8 | 4.06% | 0 | 0.00% |
| Safety | 27 | 11.63% | 4 | 14.81% | 7 | 8.33% | 1 | 14.29% |
| Non-Coding RNA | 26 | 19.36% | 3 | 11.54% | 5 | 41.34% | 2 | 40.00% |
| Informatics | 25 | 7.16% | 1 | 4.00% | 4 | 12.50% | 1 | 25.00% |
| Particles | 25 | 19.59% | 6 | 24.00% | 8 | 12.26% | 1 | 12.50% |
| Quaternary | 24 | 28.02% | 8 | 33.33% | 5 | 31.56% | 2 | 40.00% |
| Econometrics | 23 | 7.61% | 1 | 4.35% | --- | --- | --- | --- |
| Physics | 22 | 12.18% | 1 | 4.55% | 4 | 9.95% | 0 | 0.00% |
| Proteomes | 22 | 18.09% | 5 | 22.73% | 2 | 35.00% | 1 | 50.00% |
| Publications | 22 | 4.01% | 1 | 4.55% | 1 | 0.00% | 0 | 0.00% |
| Smart Cities | 22 | 8.24% | 1 | 4.55% | 7 | 6.24% | 0 | 0.00% |
| Vision | 22 | 11.46% | 1 | 4.55% | 5 | 17.32% | 0 | 0.00% |
| Acoustics | 21 | 10.69% | 2 | 9.52% | 5 | 9.84% | 0 | 0.00% |

| | All SIs, 2015-25 | | | | 2024-25 only | | | |
|---|---|---|---|---|---|---|---|---|
| | N | Mean Endogeny | N hacked SI | SI-hacking rate | N | Mean Endogeny | N hacked SI | SI-hacking rate |
| AgriEngineering | 21 | 21.93% | 6 | 28.57% | 8 | 2.96% | 0 | 0.00% |
| Medical Sciences | 21 | 15.87% | 3 | 14.29% | 1 | 0.00% | 0 | 0.00% |
| Applied System Innovation | 20 | 13.12% | 5 | 25.00% | 4 | 12.50% | 1 | 25.00% |
| Audiology Research | 20 | 22.53% | 5 | 25.00% | 9 | 17.83% | 1 | 11.11% |
| Prosthesis | 20 | 15.18% | 3 | 15.00% | 10 | 9.69% | 1 | 10.00% |
| Quantum Beam Science | 18 | 11.49% | 1 | 5.56% | 1 | 0.00% | 0 | 0.00% |
| Vehicles | 18 | 8.42% | 1 | 5.56% | 10 | 4.59% | 0 | 0.00% |
| Eng | 17 | 8.35% | 1 | 5.88% | 9 | 4.26% | 0 | 0.00% |
| Neurology International | 17 | 13.17% | 2 | 11.76% | 8 | 12.15% | 0 | 0.00% |
| Cryptography | 16 | 16.46% | 2 | 12.50% | 4 | 31.03% | 1 | 25.00% |
| Forecasting | 16 | 15.52% | 2 | 12.50% | 4 | 6.25% | 0 | 0.00% |
| Journal of Zoological and Botanical Gardens | 16 | 8.71% | 1 | 6.25% | 7 | 4.29% | 0 | 0.00% |
| Nursing Reports | 16 | 16.29% | 2 | 12.50% | 13 | 17.70% | 2 | 15.38% |
| Architecture | 15 | 9.04% | 1 | 6.67% | 11 | 7.78% | 0 | 0.00% |
| Clean Technologies | 15 | 7.92% | 2 | 13.33% | 1 | 0.00% | 0 | 0.00% |
| Epigenomes | 15 | 11.83% | 1 | 6.67% | 5 | 26.50% | 1 | 20.00% |
| Methods and Protocols | 15 | 13.89% | 2 | 13.33% | 4 | 14.18% | 1 | 25.00% |
| Tomography | 15 | 26.21% | 5 | 33.33% | 4 | 24.52% | 1 | 25.00% |
| Applied Microbiology | 13 | 5.46% | 1 | 7.69% | 8 | 7.31% | 1 | 12.50% |
| Corrosion and Materials Degradation | 12 | 10.48% | 1 | 8.33% | 5 | 2.50% | 0 | 0.00% |
| Endocrines | 12 | 11.85% | 1 | 8.33% | 3 | 6.67% | 0 | 0.00% |
| Plasma | 12 | 19.92% | 3 | 25.00% | 3 | 8.33% | 0 | 0.00% |
| Challenges | 11 | 22.99% | 3 | 27.27% | --- | --- | --- | --- |
| Diabetology | 11 | 12.40% | 1 | 9.09% | 5 | 16.66% | 1 | 20.00% |
| Literature | 11 | 15.76% | 2 | 18.18% | 5 | 17.34% | 1 | 20.00% |
| Surfaces | 11 | 13.12% | 2 | 18.18% | 4 | 7.78% | 0 | 0.00% |
| CivilEng | 10 | 18.11% | 1 | 10.00% | 3 | 21.10% | 0 | 0.00% |
| Fuels | 10 | 10.71% | 1 | 10.00% | 6 | 5.00% | 0 | 0.00% |
| Gastroenterology Insights | 9 | 12.90% | 1 | 11.11% | 5 | 2.22% | 0 | 0.00% |
| Journal of Cybersecurity and Privacy | 9 | 15.92% | 1 | 11.11% | 5 | 22.00% | 1 | 20.00% |
| Quantum Reports | 9 | 15.94% | 1 | 11.11% | 3 | 16.67% | 1 | 33.33% |
| Electricity | 8 | 10.12% | 1 | 12.50% | 5 | 12.86% | 1 | 20.00% |
| European Burn Journal | 8 | 19.66% | 2 | 25.00% | 4 | 17.50% | 1 | 25.00% |
| Hemato | 8 | 21.80% | 2 | 25.00% | 1 | 0.00% | 0 | 0.00% |
| Mining | 8 | 16.68% | 1 | 12.50% | 6 | 20.57% | 1 | 16.67% |

| | All SIs, 2015-25 | | | | 2024-25 only | | | |
|---|---|---|---|---|---|---|---|---|
| | N | Mean Endogeny | N hacked SI | SI-hacking rate | N | Mean Endogeny | N hacked SI | SI-hacking rate |
| Signals | 8 | 10.41% | 1 | 12.50% | 2 | 0.00% | 0 | 0.00% |
| Stresses | 8 | 21.35% | 2 | 25.00% | --- | --- | --- | --- |
| Tourism and Hospitality | 8 | 11.66% | 1 | 12.50% | 5 | 12.00% | 1 | 20.00% |
| Transplantology | 8 | 27.65% | 3 | 37.50% | --- | --- | --- | --- |
| Biomechanics | 7 | 14.73% | 1 | 14.29% | 5 | 17.28% | 1 | 20.00% |
| Ecologies | 7 | 15.51% | 2 | 28.57% | 2 | 10.55% | 0 | 0.00% |
| Forensic Sciences | 7 | 15.59% | 1 | 14.29% | 4 | 25.00% | 1 | 25.00% |
| Liquids | 7 | 24.43% | 1 | 14.29% | 5 | 18.96% | 0 | 0.00% |
| Sexes | 7 | 13.24% | 1 | 14.29% | 3 | 4.17% | 0 | 0.00% |
| Uro | 7 | 24.36% | 2 | 28.57% | 3 | 7.40% | 0 | 0.00% |
| Epidemiologia | 6 | 20.35% | 1 | 16.67% | 1 | 0.00% | 0 | 0.00% |
| Foundations | 6 | 38.33% | 2 | 33.33% | --- | --- | --- | --- |
| Hearts | 6 | 37.88% | 3 | 50.00% | --- | --- | --- | --- |
| Journal of Molecular Pathology | 6 | 22.12% | 2 | 33.33% | 2 | 0.00% | 0 | 0.00% |
| BioTech | 5 | 13.34% | 1 | 20.00% | 1 | 0.00% | 0 | 0.00% |
| Dermatopathology | 5 | 31.44% | 2 | 40.00% | 2 | 50.00% | 1 | 50.00% |
| Metrology | 5 | 8.00% | 1 | 20.00% | 3 | 0.00% | 0 | 0.00% |
| Pathophysiology | 5 | 28.66% | 2 | 40.00% | --- | --- | --- | --- |
| Reproductive Medicine | 5 | 16.00% | 1 | 20.00% | 1 | 0.00% | 0 | 0.00% |
| Scientia Pharmaceutica | 5 | 20.78% | 2 | 40.00% | 1 | 4.30% | 0 | 0.00% |
| Clinical and Translational Neuroscience | 4 | 30.55% | 1 | 25.00% | 2 | 21.10% | 0 | 0.00% |
| Dairy | 4 | 43.85% | 3 | 75.00% | --- | --- | --- | --- |
| Electronic Materials | 4 | 15.00% | 1 | 25.00% | 1 | 0.00% | 0 | 0.00% |
| Oxygen | 4 | 30.20% | 1 | 25.00% | 1 | 100.00% | 1 | 100.00% |
| Pollutants | 4 | 15.00% | 1 | 25.00% | 2 | 0.00% | 0 | 0.00% |
| Analytics | 3 | 38.10% | 1 | 33.33% | 2 | 57.15% | 1 | 50.00% |
| Electrochem | 3 | 22.23% | 1 | 33.33% | 1 | 50.00% | 1 | 100.00% |
| Hematology Reports | 3 | 70.00% | 2 | 66.67% | 1 | 100.00% | 1 | 100.00% |
| Humans | 3 | 26.87% | 1 | 33.33% | 1 | 12.50% | 0 | 0.00% |
| Immuno | 3 | 60.33% | 2 | 66.67% | --- | --- | --- | --- |
| Knowledge | 3 | 16.67% | 1 | 33.33% | 3 | 16.67% | 1 | 33.33% |
| Network | 3 | 40.00% | 1 | 33.33% | --- | --- | --- | --- |
| Seeds | 3 | 25.57% | 1 | 33.33% | 1 | 20.00% | 0 | 0.00% |

| | All SIs, 2015-25 | | | | 2024-25 only | | | |
|---|---|---|---|---|---|---|---|---|
| | N | Mean Endogeny | N hacked SI | SI-hacking rate | N | Mean Endogeny | N hacked SI | SI-hacking rate |
| GeoHazards | 2 | 55.00% | 2 | 100.00% | --- | --- | --- | --- |
| Telecom | 2 | 21.45% | 1 | 50.00% | --- | --- | --- | --- |
| Thalassemia Reports | 2 | 21.45% | 1 | 50.00% | --- | --- | --- | --- |